\documentclass[pra,superscriptaddress,twocolumn,floatfig,showpacs,floatfix]{revtex4-1}

\usepackage{graphicx,color}
\usepackage{amsmath,amssymb,bm}
\usepackage{braket}		% Dirac notation

\usepackage{mathtools}
\usepackage{float}
\usepackage[plainpages=false,pdfpagelabels,colorlinks=true,linkcolor=blue,urlcolor=blue,citecolor=blue,pdftitle={Finite-temperature density matrix renormalization group for electron-phonon systems and Holstein polaron spectral functions},pdfauthor={},pdfdisplaydoctitle=true,pdfduplex=DuplexFlipLongEdge]{hyperref}

\definecolor{darkred}{rgb}{0.90,0,0}
\definecolor{darkgreen}{rgb}{0,0.60,.2}
\definecolor{darkblue}{rgb}{0,0,1}
\definecolor{grey}{cmyk}{0,0,0,0.25}
\definecolor{orange}{cmyk}{0,0.6,1,0}

\usepackage{physics}
\usepackage{mathrsfs}
\begin{document}
\title{
Finite-temperature density-matrix renormalization group  method for electron-phonon systems: Thermodynamics and Holstein-polaron spectral functions
}

\author{David Jansen}
\affiliation{Institut f\"ur Theoretische Physik, Georg-August-Universit\"at G\"ottingen, D-37077 G\"ottingen, Germany}

\author{Janez Bon\v{c}a}
\affiliation{J. Stefan Institute, 1000 Ljubljana, Slovenia}
\affiliation{Faculty of Mathematics and Physics, University of Ljubljana, 1000 Ljubljana, Slovenia}

\author{Fabian Heidrich-Meisner}
\affiliation{Institut f\"ur Theoretische Physik, Georg-August-Universit\"at G\"ottingen, D-37077 G\"ottingen, Germany}

\begin{abstract}
We investigate the thermodynamics and finite-temperature spectral functions of the Holstein polaron using a density-matrix renormalization group method. Our method combines purification and local basis optimization (LBO) as an efficient treatment of phonon modes. LBO is a scheme which relies on finding the optimal local basis by diagonalizing the local reduced density matrix. By transforming the state into this basis, one can truncate the local Hilbert space with a negligible loss of accuracy for a wide range of parameters. In this work, we focus on the crossover regime between large and small polarons of the Holstein model. Here, no analytical solution exists and we show that the thermal expectation values at low temperatures are independent of the phonon Hilbert space truncation provided the basis is chosen large enough. We then demonstrate that we can extract the electron spectral function and establish consistency with results from a finite-temperature Lanczos method. We additionally calculate the electron emission spectrum and the phonon spectral function and show that all the computations are significantly simplified by the local basis optimization. We observe that the electron emission spectrum shifts spectral weight to both lower frequencies and larger momenta as the temperature is increased. The phonon spectral function experiences a large broadening and the polaron peak at large momenta gets significantly flattened and merges almost completely into the free-phonon peak. 
\end{abstract}

\maketitle
\section{Introduction}
\label{sec:intro}

Developments in experimental methods with ultra-fast dynamics (see, e.g., Refs.~\cite{gadermaier10,novelli14,dalconte15,Giannetti_capone_16,hwang_2019} and Refs.~\cite{basov_11,orenstein12} for a review) have reinforced the interest in the theoretical modeling of electron-phonon interactions. Despite the complexity of real materials, qualitative insights can be gained from model systems such as the Holstein model~\cite{Holstein1959,hirsch_83,wellein96,wellein97,wellein98,bonca99,zhang99,weisse00,fehske_2000,
weisse02,ku2002,hohenadler_03,hohenadler04,hohenadler_05,fehske2007,ku2007,barisic08,
ejima09,alvermann10,vidmar10,vidmar11,goodvin11,defilippis12b,
sayyad14,dorfner_vidmar_15,huang_17,dutreix_17,
brockt_17,kemper_17,stolpp2020}, Hubbard-Holstein model~\cite{fehske04,werner07,fehske08,golez12a,defilippis12,werner13,
nocera14,werner_eckstein_15,sentef_16,hashimoto_ishihara_17},
hard-core boson-phonon models~\cite{dey_15,kogoj_mierzejewski_16},
 the t-J model augmented with phonons~\cite{vidmar09,vidmar11c} and spin-boson models~\cite{Guo2012,bruognolo_14,wall_16,sato_18}. It is believed that many experimental results can be interpreted by studying such toy Hamiltonians that  contain important key features. A paradigmatic example is the Holstein-polaron model \cite{Holstein1959} which consists of one electron interacting with local bosons. Since the electron-boson interaction is the only mean of thermalization in the system (see, e.g., Ref.~\cite{jansen19}), the model allows us to study this particular relaxation channel, which plays an important role in real materials, in a controlled way.

The Holstein polaron at zero temperature has been the subject of intense research \cite{bonca99,ku2002,barisic02,barisic04,barisic06,loos_2006,loos_hohenadler_2006,vidmar10}. Recent developments in the field of thermalization in isolated quantum systems and quench dynamics have fueled the demand for additional research on the model at a finite temperature \cite{demello97,Paganelli_2006,rigol_dunjko_08,rigol_srednicki_12,sorg14,mischenko_15,dalessio_kafri_16,lipeng_17,deutsch_18,mori_ikeda_18,jansen19}. In particular, the momentum dependence of the spectral function and the self-energy have recently been computed using a finite-temperature Lanczos method by
 Bon\v{c}a \textit{et al.} in Ref.~\cite{bonca2019}, followed by a comparison between spectral properties of the Holstein polaron and an electron coupled to  hard-core bosons~\cite{bonca2020}.

In this paper, we introduce an alternative numerical approach. We demonstrate that a density-matrix renormalization group (DMRG) method~\cite{white92,schollwock2005density,schollwock2011density} can  efficiently reproduce the spectral function. We also compute the electron emission spectrum, which can be accessed in angle-resolved photoemission spectroscopy (ARPES) experiments~\cite{eberhardt_80,damascell_03,damascelli_04,Kirkegaard_2005,freericks_09,Hofmann_2009}, and the phonon spectral function. The method further allows us to compute thermodynamic observables for very large system sizes compared to other wave-function based methods.

We combine finite-temperature DMRG with purification~\cite{verstrate2004,feiguin2005,barthel2009,feiguin2010}, time-dependent DMRG (tDMRG)~\cite{daley2004,white04,vidal2004,schollwock2011density,paeckel_2019} and local basis optimization (LBO)~\cite{zhang98} to obtain an efficient scheme to both generate the finite-temperature matrix-product state (MPS) and to compute different Green's functions. LBO, originally introduced by Zhang \textit{et al.} in Ref.~\cite{zhang98}, has already been used for the real-time evolution~\cite{brockt_dorfner_15,shroeder_16,brockt_17,stolpp2020} and ground-state algorithms ~\cite{bursil_99,Friedman2000,barford_02,barford_06,wong_08,Guo2012,tozer_14,dorfner_FHM_16,stolpp2020}. The DMRG method with purification requires an infinite-temperature state as its starting point which is artificial and strongly dependent on the truncation of the phonon Hilbert space. Our results, however, become independent of that truncation in the polaron-crossover regime at low temperatures, which is physically most relevant.
This allows for the efficient computation of static and dynamic properties of the Holstein model at finite temperatures.

In particular, using finite-temperature states we compute the electron addition spectral function. This quantity has already been analyzed thoroughly at finite temperatures by Bon\v{c}a \textit{et al.} in Ref.~\cite{bonca2019} for a system with periodic boundary conditions. We show that we can resolve the same peaks as the finite-temperature Lanczos method and observe an excellent quantitative agreement. We additionally compute the electron emission spectrum and the phonon spectral function. The electron emission spectrum was computed in Ref.~\cite{demello97} for a two-site and two-electron system. Here, we focus on one electron and go up to twenty one sites.  We further show that the LBO scheme proposed in Ref.~\cite{brockt_dorfner_15} becomes computationally beneficial at low temperatures and significantly simplifies the computations of these spectral functions.

This paper is structured as follows. In Sec.~\ref{sec:model}, we introduce the Holstein-polaron model, the observables, and the spectral functions. We proceed in Sec.~\ref{sec:method} with a description of the methods used. We introduce DMRG with purification in Sec.~\ref{subsec:mps}, the time-evolution algorithm in Sec.~\ref{subsec:lbo} and the finite-temperature Lanczos method in Sec.~\ref{subsec:ftlm}.  In Sec.~\ref{subsec:ssAp}  and Sec.~\ref{subsec:ssD}, we show the spectral functions for the single-site Holstein model.  We present the results for the thermodynamic expectation values in Sec.~\ref{sec:TD} and the results for the spectral functions in Sec.~\ref{sec:specfuncres}. In Sec.~\ref{sec:summary}, we summarize the paper and provide an outlook.
\section{Model}
\label{sec:model}
 \subsection{The Holstein polaron}
  \label{subsec:holpol}
To study finite-temperature polaron properties we consider the single-electron Holstein model~\cite{Holstein1959}. The Hamiltonian is defined as
\begin{equation} \label{eq:def_HolHam}
  \hat H =\hat H_{\rm kin} + \hat H_{\rm ph} + \hat H_{\rm e-ph}\, .
\end{equation}
The model has $L$ sites and we use open boundary conditions, unless stated otherwise. We set $\hbar=1$ throughout this paper.
The first term, the kinetic energy of the electron, then becomes
\begin{equation} \label{eq:def_Hkin}
  \hat H_{\rm kin} =- t_0 \sum_{j=1}^{L-1} \left( \hat c_j^{\dag} \hat c_{j+1}^{\phantom{\dag}} + \hat c_{j+1}^{\dag} \hat c_j^{\phantom{\dag}} \right)\, ,
\end{equation}
with $\hat c^{\dag}_j (\hat c_j)$ being the electron creation (annihilation) operator on site $j$ and $t_0$ the hopping amplitude. The second term is the phonon energy
\begin{equation} \label{eq:def_Hph}
    \hat H_{\rm ph}= \omega_0 \sum_{j=1}^L \hat b_j^{\dag}  \hat b_{j}^{\phantom{\dag}} \, ,
  \end{equation}
where $\hat b^{\dag}_j (\hat b_j)$ is the creation (annihilation) operator of an optical phonon on site $j$ with the constant frequency $\omega_0$. The last term is the electron-phonon coupling 
\begin{equation} \label{eq:def_Heph}
    \hat H_{\rm e-ph}= \gamma \sum_{j=1}^L\hat n_j \left( \hat b_j^{\dag} + \hat b_{j}^{\phantom{\dag}}  \right) \, ,
  \end{equation}
  with $\hat n_j= \hat c^{\dag}_j \hat c_j$. We furthermore define the dimensionless coupling parameter
        \begin{equation} \label{eq:def_lam}
          \lambda=\frac{\gamma^2}{2t_0 \omega_0} \, ,
        \end{equation}
        which characterizes the crossover from a large ($\lambda<1$) to a small ($ \lambda>1$) polaron. In this work, we focus  on the intermediate regime and set $\lambda=1$. For a discussion of other parameter regimes, see Appendix~\ref{sec:app1}. 
          \subsection{Thermodynamics}
  \label{subsec:td}
        We first want to study the thermodynamics of the model. The thermal expectation value of an observable $\hat O$ in the canonical ensemble at temperature $T$ is defined as
                        \begin{equation} \label{eq:def_Et}
          \expval*{\hat O}_T = \text{Tr}[\hat \rho (T) \hat O] \, ,
        \end{equation}
        where $\hat \rho(T)$ is the thermal density matrix at temperature $T$. In the canonical ensemble,
                 \begin{equation} \label{eq:def_rho_can}
          \hat{\rho}(T)= \frac{1}{Z}e^{-\beta \hat{H} }\, ,
        \end{equation}
        where we have set $k_{\rm B}=1$ such that $\beta=1/T$ and $Z$ is the partition function.
        We will focus on four observables: The total energy $E(T)=\expval*{\hat H}_T$, the kinetic energy $E_{\rm kin}(T)=\expval*{\hat H_{\rm kin}}_T$, the coupling energy $E_{\rm e-ph}(T)=\expval*{\hat H_{\rm e-ph}}_T$ and the phonon energy $E_{\rm ph}(T)=\expval*{\hat H_{\rm ph}}_T$.
                 \subsection{Spectral functions}
  \label{subsec:specfunc}
We are also interested in dynamical quantities by investigating Green's functions of operators acting on sites $m$ and $n$. 
We define the greater Green's function  
\begin{equation} \label{eq:def_corel0}
G_{T,0}^>(m,n, t)= -i \expval*{\hat c_m(t)\hat c^{\dagger}_{n}(0)}_{T,0} \, ,
\end{equation}
where the sub-indices $T,0$ indicate that the thermal expectation value is calculated in the zero-electron sector. Since we use open boundary conditions, we construct the Fourier transform into quasi-momentum space (see, e.g., Refs.~\cite{Ejima_2011,benthien_04}) as
\begin{equation} \label{eq:def_corel0FT}
\hat c_k = \sqrt{\frac{2}{L+1}} \sum\limits_{j=1}^L \sin(k j)\hat c_j  \, ,
\end{equation}
where $k=\pi m_k /(L+1)$ and $1\leq m_k \leq L$.
The greater Green's function in $k$ and $\omega$ space then becomes
 \begin{equation} \label{eq:def_great}
G^{>}_{T,0}( k, \omega)=  -i \int_{-\infty}^{\infty} dt e^{i \omega t -\abs{t}\eta}G_{T,0}(k, t) \, ,
\end{equation}
where $\eta =0^+$ is an artificial broadening. From the greater Green's function, we extract the electron spectral function
 \begin{equation} \label{eq:def_ferspec}
A(k, \omega)=-\frac{1}{2 \pi} \text{Im}[G^{>}_{T,0}(k,\omega)] \, .
\end{equation}
Since we are in the zero-electron sector, Eq.~\eqref{eq:def_ferspec} contains all the information about the spectrum.
%This spectral function .
Here, we extend previous studies \cite{bonca2019,demello97} of the finite-temperature Holstein polaron by also computing the lesser Green's function in the one-electron sector
 \begin{equation} \label{eq:def_corel1}
G^<_{T,1}( m,n, t)=i \expval*{\hat c^{\dagger}_{m}(0) \hat c_n(t)}_{T,1} \, ,
\end{equation} which we use to obtain the electron emission spectrum
   \begin{equation} \label{eq:def_spectild}
A^+(k, \omega)=-\frac{1}{2 \pi}\text{Im}[ -G^{<}_{T,1}( k, \omega)] \, .
\end{equation}
We are further interested in the greater Green's function in the phonon sector
 \begin{equation} \label{eq:def_corelD}
D^{>}_{T,1}( m,n, t)= -i\expval*{\hat X_{m}(t) \hat X_n(0)}_{T,1} \, ,
\end{equation} 
 where $\hat X_{n}=\hat b_n + \hat b_{n}^{\dagger}$ is the phonon displacement. We use Eq.~\eqref{eq:def_corelD} to calculate
 the phonon spectral function
   \begin{equation} \label{eq:def_phospec}
B(k, \omega)=-\frac{1}{2 \pi} \text{Im}[D^>_{T,1}( k, \omega)] \, . 
\end{equation}
\section{Methods}
In this section, we first describe our main numerical method, DMRG using purification and local basis optimization.
We next briefly review the finite-temperature Lanczos method used in Ref.~\cite{bonca2019}.
In order to guide the discussion of the numerical results, we compute the three spectral functions in the single-site limit.
\label{sec:method}
  \subsection{Density-matrix renormalization group with purification}
  \label{subsec:mps}
  While the density-matrix renormalization group and matrix-product states were originally developed to find ground states~\cite{white92,schollwock2005density,schollwock2011density}, they have proven to be extremely useful tools for calculating spectral functions~\cite{tiegel_14,hallberg_95,kuener_99,jeckelmann_02,holzner_11,weichselbaum_09},  carrying out the  time evolution~\cite{daley2004,white04,vidal2004,schollwock2011density,paeckel_2019}, and finite-temperature calculations~\cite{verstrate2004,zwolak2004,sirker2005,feiguin2005,barthel2009,karrasch2012,barthel2012,barthel2013,karrasch2013,kennes2016,hauschild2018} as well.
  There are several ways to utilize matrix-product states at finite temperatures. These include, among others,  minimally entangled typical thermal state algorithms~\cite{white2009,stoudenmire2010,Bruognolo_17}, purification algorithms~\cite{feiguin2005,barthel2009,feiguin2010,nocera_16} or a mixture of both~\cite{chung19,chen_stoudemir2019}.
  
  In this paper, we use the purification method \cite{verstrate2004} which doubles the system by adding an auxiliary space to the physical Hilbert space $\mathscr{H}_{\rm P} \rightarrow \mathscr{H}_{\rm P} \otimes  \mathscr{H}_{\rm A}$. One can write down a state in this doubled Hilbert space $\ket{\psi} \in \mathscr{H}_{\rm P} \otimes  \mathscr{H}_{\rm A} $. If one now traces out the auxiliary space, one can simulate a mixed state in the physical Hilbert space with the density matrix  
  \begin{equation}
  \label{eq:redDM}
  \hat{\rho}_{\rm P}=\text{Tr}_{\rm{A}} [\ket{\psi} \bra{\psi}] \, .
  \end{equation} 
 Since we are working with a fixed number of electrons, we will use the notation where $\ket*{\psi_{\beta}^n}$ represents a state at temperature $T$ with $n$ electrons in the physical system. For example, the state $\ket*{\psi_{\beta=0}^1}\in \mathscr{H}_P \otimes  \mathscr{H}_A $ can be expressed analytically. This state can then be used to simulate the density matrix from Eq.~\eqref{eq:def_rho_can} for one electron at $\beta=0$. The density matrix for $\beta \neq 0$ is generated by the evolution in imaginary time of $\ket*{\psi_{\beta=0}^n} $ as $\ket*{\psi_{\beta}^n}=e^{-\hat H\beta /2}\ket*{\psi_{\beta=0}^n} $.  The thermal expectation value of an observable $\hat{O}$ can then be calculated as
    \begin{equation}
  \label{eq:mpsexp}
  \expval*{\hat{O}}_{T,n} = \frac{\expval*{\hat{O}}{\psi^n_{\beta}}}{\ip*{\psi^n_{\beta}}}\, .
\end{equation}

  Since we use thermal states with both one and zero electrons we briefly illustrate how to generate both of them at $\beta=0$. Since the Holstein model contains infinitely many local phonon degrees of freedom, we first introduce a local cutoff $M$ which represents the maximal number of phonons on each site. We furthermore define a local basis state  $\ket{\sigma_i}=\ket*{n^{\rm e}_i, n^{\rm ph}_i}$ with electron occupation $n_i^{\rm e} \in \{ 0,1 \}$ and phonon occupation $n_i^{\rm ph} \in \{ 0,\hdots M \}$. This determines the local dimension $d=2(M+1)$. We further write $\ket*{\vec{\sigma}}= \ket*{\sigma_1, \sigma_2, \hdots ,\sigma_L}$. The zero-electron state $\ket*{\psi^0_{\beta=0}}$ becomes
  \begin{equation} \label{eq:MPS1}
     \ket*{\psi^0_{\beta=0}}=\sum\limits_{\vec{\sigma}^{\rm P}, \vec{\sigma}^{\rm A}}A^{\sigma_1^{\rm P},\sigma_1^{\rm A}}_1 A^{\sigma_2^{\rm P}, \sigma_2^{\rm A}}_2\hdots A^{\sigma_L^{\rm P}, \sigma_L^{\rm A}}_L\ket*{\vec{\sigma}^{\rm P},\vec{\sigma}^{\rm A}}\, ,
  \end{equation}
where each $A_i^{\sigma_i^{\rm{P}}, \sigma_i^{\rm{A}}}= \delta_{n_{i}^{\rm{e,A}},0}\delta_{n_{i}^{e,P},0}\delta_{n^{\rm ph,P}_{i},n^{ \rm ph,A}_{i}}$ is the local tensor corresponding to maximum entanglement between the physical site $\sigma_i^{{\rm P}}$ and the auxiliary site $\sigma_i^{\rm A}$. To generate the one-electron state $ \ket*{\psi^1_{\beta=0}}$, we proceed in a similar fashion as in Ref.~\cite{barthel_16}. We first write down the maximum entangled one-electron tensor  $\tilde{A}_j^{\sigma_{j}^{\rm P}, \sigma_{j}^{\rm A}}=\delta_{n_{j}^{e, \text{A}},1}\delta_{n_{j}^{e, \text{P}},1}\delta_{n^{\rm ph, A}_{ j},n^{\rm  ph, P}_{j}}$. We then define  our wave function as the superposition of terms which all have the one-electron tensor $\tilde{A}_j^{\sigma_{j}^{\rm P}, \sigma_{j}^{\rm A}}$ at a different site $j$. On the sites $i\neq j$, we just place the zero-electron tensors $A_i^{\sigma_i^{\rm{P}}, \sigma_i^{\rm{A}}}$ from Eq.~\eqref{eq:MPS1}. The total wave function becomes
    \begin{equation} \label{eq:MPS2}
      \ket*{\psi^1_{\beta=0}}=\sum\limits_{j=1}^L\sum\limits_{\vec{\sigma}^{\rm P}, \vec{\sigma}^{\rm A}}A^{\sigma_1,\sigma_1^{\prime}}_1 \hdots \tilde{ A}^{\sigma_j, \sigma_j^{\prime}}_{j}\hdots A^{\sigma_L, \sigma_L^{\prime}}_L\ket*{\vec{\sigma}^{\rm P},\vec{\sigma}^{\rm{A}}}\, .
    \end{equation}
   After constructing $\ket*{\psi^n_{\beta=0}}$, one then generates the desired  $\ket*{\psi^n_{\beta}}$ by imaginary-time evolution.
  \subsection{Time evolution with local basis optimization}
  \label{subsec:lbo}
Since we are interested in thermodynamics and spectral functions at finite temperatures, we need to carry out both imaginary and real-time evolution. This subsection explains the procedure and its application to electron-phonon systems. 
To calculate the time evolution we use tDMRG~\cite{daley2004,white04} combined with local basis optimization (LBO)~\cite{zhang98}. The idea of the local basis optimization is to find a numerically efficient representation of the phonon Hilbert space, thus reducing the computational cost. This has already been combined with exact diagonalization to study zero-temperature dynamical properties of the Holstein model, e.g., by Zhang \textit{et al.} in Ref.~\cite{zhang99}, and with matrix-product states to calculate the real-time evolution of pure states, e.g., by Brockt \textit{et al.}~\cite{brockt_dorfner_15} for the polaron problem, by Stolpp \textit{et al.}~\cite{stolpp2020} for a charge density wave, as well as in Refs.~\cite{shroeder_16,brockt_17}. In this paper, we demonstrate that the local basis optimization is computationally beneficial for computing thermodynamics and real-time evolution at finite temperature for the Holstein polaron.

  For the tDMRG method, we first write our Hamiltonian as a sum of terms  $\hat h_{l}$  which act on the two neighboring sites $l$ and $l+1$. For a time step $dt$ ($-id \tau $ for imaginary-time evolution) one can then carry out  a second-order Trotter-Suzuki decomposition into even and odd terms
     \begin{equation} \label{eq:trot}
e^{-idt\hat H}=e^{-idt  \hat H_{\rm even} /2}e^{-idt \hat H_{\rm odd}}e^{-idt \hat H_{\rm even} /2}+ O(dt^3)\, .
\end{equation}
One can further write each exponential as the product of local elements $e^{-idt  \hat H_{\rm even} /2}=\prod\limits_{l: \rm even} e^{-idt  \hat h_{l} /2}$. Since each physical site is connected to an auxiliary site, one must first apply a fermionic swap gate~\cite{stoudenmire2010,orus_14} to swap site $\sigma_{l}^{P}$ and $\sigma^{\rm A}_{l}$. One then acts with the time evolution gate $e^{-idt  \hat h_{l} /2}$ followed by another gate which swaps $\sigma_{l}^{\rm P}$ and $\sigma^{\rm A}_{l}$ back.

To illustrate how the local basis optimization works, we assume that our MPS is already in an optimal local basis $\ket*{\mathbf{\tilde{\sigma}^{\rm P}}}$ and that the local transformation matrices $R^{\tilde{\sigma}^{\rm P}_i}_{\sigma_i^{\rm P}}$ transform the physical index from the bare into the optimal basis. We then first transform two of the legs from our time-evolution gate $U_{\sigma_i^{\rm P}, \sigma_{i+1}^{\rm P}}^{ \sigma_i^{\prime \rm P}, \sigma_{i+1}^{\prime \rm P}} $ to
     \begin{equation} \label{eq:trafgate}
U_{\tilde{\sigma}^{\rm P}_i, \tilde{\sigma}^{\rm P}_{i+1}}^{\sigma_i^{\prime \rm P}, \sigma^{\prime \rm P}_{i+1} }= R^{\sigma_i^{\rm P}}_{\tilde{\sigma}^{\rm P}_i}R^{\sigma^{\rm P}_{i+1}}_{\tilde{\sigma}^{\rm P}_{i+1}}U_{\sigma^{\rm P}_i, \sigma^{\rm P}_{i+1}}^{ \sigma^{\prime \rm P}_i, \sigma_{i+1}^{\prime \rm P}}\, .
\end{equation}
We then apply this gate to the two-site tensor $ M^{\tilde{\sigma}^{\rm P}_i, \tilde{\sigma}^{\rm P}_{i+1}}=A^{\tilde{\sigma}^{\rm P}_i}A^{ \tilde{\sigma}^{\rm P}_{i+1}}$ and get  
     \begin{equation} \label{eq:applygate}
\phi^{\sigma_i^{\prime \rm P}, \sigma_{i+1}^{\prime  \rm P}}=U_{\tilde{\sigma}^{\rm P}_i, \tilde{\sigma}^{\rm P}_{i+1}}^{ \sigma_i^{\prime \rm P}, \sigma_{i+1}^{\prime \rm P}} M^{\tilde{\sigma}^{\rm P}_i, \tilde{\sigma}^{\rm P}_{i+1}}\, .
\end{equation}
We then generate the local reduced density matrix
     \begin{equation} \label{eq:rho}
\rho_{\sigma_i^{\rm P}}^{ \sigma_{i}^{\prime \rm P}}=\phi^{\dagger }_{\sigma_i^{\rm P}, \sigma_{i+1}^{\rm P}}\phi^{\sigma_i^{\prime \rm P}, \sigma_{i+1}^{\rm P}}\, ,
\end{equation}
which we diagonalize such that
  \begin{equation} \label{eq:rho_diag}
\rho_{\sigma_i^{\rm P}}^{ \sigma_{i}^{\prime \rm P}}=R_{\sigma_i^{\rm P}}^{\tilde{\sigma}^{\rm P}_i}D^{\tilde{\sigma}_i^{\prime \rm P}}_{\tilde{\sigma}^{\rm P}_i }R_{ \tilde{\sigma}^{\prime  \rm P}_i}^{ \dagger\sigma_i^{\prime \rm P}} \, .
\end{equation}
The matrix $ R_{\sigma_i^{\rm P}}^{\tilde{\sigma}_i^{\rm P}}$ is now the updated transformation matrix which rotates the site $i$ into the adapted optimal basis. The transformation matrices can then be applied to $\phi^{\sigma_i^{\prime}, \sigma_{i+1}^{\prime}}$ from Eq.~\eqref{eq:applygate} before the following singular value decomposition. If the optimal basis has dimension $d_{\rm LBO}$ and the MPS has a bond dimension $\chi$, the cost of the SVD has then changed from $O(d^3\chi^3) $ to $O(d_{\rm LBO}^3\chi^3)$ \cite{brockt_dorfner_15}. The transformation is only beneficial if we can truncate the optimal basis such that $d_{\rm LBO}\ll d$, since the transformation itself has a cost of $O(d^3\chi^2)$ for building the reduced-density matrix, $O(d^3)$ for the diagonalization thereof and $O(d^2d_{\rm LBO}\chi^2)$ for the basis transformation~\cite{brockt_dorfner_15}. To control the truncation, we discard the smallest eigenvalues $w_{\alpha}$ such that the truncation error is below a threshold: $\sum\limits_{\alpha \in \rm{discarded} } w_{\alpha}/(\sum\limits_{\rm{all} \: \alpha} w_{\alpha})<\rho_{\rm LBO}$. When carrying out a regular singular-value decomposition in the tDMRG algorithm, we discard all singular values such that $\sum\limits_{\alpha \in \rm{discarded }} s_{\alpha}^2/(\sum\limits_{\rm{all} \:  \alpha } s_{\alpha}^2)<\rho_{\rm bond}$. 

To calculate a general correlation function $C_{T,n}(\omega, k)$, such as the Green's functions in Eqs.~\eqref{eq:def_corel0},~\eqref{eq:def_corel1} and~\eqref{eq:def_corelD}, we first obtain the desired state $\ket*{\psi^n_{\beta}} $ through imaginary-time evolution. In that process, we apply the LBO only  to the physical sites such that only these are in their optimal basis. Before we start the real-time evolution, we first iterate through the MPS and obtain the optimal basis for both the physical and the auxiliary sites by creating the matrix $M^{\sigma_i^{\rm P}, \sigma^{\rm A}_{i}}$  and getting the transformation matrices $ R_{\sigma_i^{\rm P}}^{\tilde{\sigma}^{\rm P}_i}$ and  $ R_{\sigma^{\rm A}_i}^{\tilde{\sigma}^{\rm A}_i}$. We then follow Ref.~\cite{barthel2009} and compute the desired correlation functions 
\begin{equation} \label{eq:corel}
C_{T,n}(m,l, t)=\expval*{\hat A_m(t/2)\hat B_l(-t/2)}{\psi^n_{\beta}} \, ,
\end{equation} where $\hat A, \hat B$ are general operators and\begin{equation} \label{eq:optev}
\hat A_m(t)=\hat U^{\rm P \dagger }(t/2)\hat U^{\rm A \dagger }(t/2)\hat A_m \hat U^{\rm P}(t/2)U^{\rm A}(t/2)\, .
\end{equation}
In Eq.~\eqref{eq:optev}, $\hat U^{\rm P}(t/2)=e^{-i \hat H^{\rm P} t/2}$ acts on the part of the state in the physical Hilbert space and  $\hat U^{\rm A}(t/2)=e^{i \hat  H^{\rm A} t/2}$ correspondingly time evolves the part of the state in the auxiliary space in the opposite direction. This is done to keep the entanglement entropy low during the real-time evolution~\cite{barthel2013,barthel2012,karrasch2013,karrasch2012,kennes2016}. Even though this procedure is not optimal~\cite{hauschild2018}, it provides a natural extension of the local basis optimization to the auxiliary sites and allows us to keep them in an optimal basis during the time evolution.  
We then obtain $C_{T,n}(k, \omega)$ from $C_{T,n}( m,l, t)$ by Fourier transformations in space and time.

We further use linear prediction \cite{barthel2009,white_affleck2008,vaidyanathan_08} to access larger times. Since we do the real-time evolution on states $\ket*{\phi^n_{\beta}}=\hat A_l \ket*{\psi^{n}_{\beta}}$, which are not normalized, e.g., $\ip*{\phi^n_{\beta}} \neq 1$, we do not re-normalize the state after applying the time-evolution gate. We observe that the norm of $\ket*{\phi^n_{\beta}}$ is still of $O(1)$, so that we can apply the same truncation criteria as if we worked with normalized states. For all imaginary-time evolutions, we use $d \tau  \omega_0=0.1$ and for the real-time evolution, we use $d t  \omega_0=0.01$. The real-time evolution is done up to a maximum time $t_{\rm max} \omega_0$. The accessible $t_{\rm max} \omega_0$ depends on the observable and model parameters and is determined by the computational resources available. All DMRG calculations are carried out using~\cite{ITensor}. 
 \subsection{Finite-temperature Lanczos method}
  \label{subsec:ftlm}
  We now proceed by introducing alternative methods used as benchmarks.  The thermodynamic quantities will be compared to exact diagonalization (ED) and the spectral function from Eq.~\eqref{eq:def_ferspec} will be compared to the finite-temperature Lanczos method~\cite{jaklic2000,Prelovsek2013} (FTLM).
 
 The finite-temperature Lanczos method data used here is obtained from Ref.~\cite{bonca2019}. The electron spectral function is expressed as
    \begin{multline} \label{eq:AFTLM}
      A( k,\omega)= Z^{-1} \sum\limits_{r=1}^R \sum\limits_{j=1}^M \sum\limits_{n=1}^Ne^{- \beta \epsilon _n^0} \bra{r^0} \ket{\phi ^0_n}\matrixel{\phi ^0_n}{\hat c_k}{\psi_j} \\
     \times \matrixel{\psi_j}{\hat c_k^{\dagger}}{r^0} \delta(\omega- \epsilon_j + \epsilon_n^0) \, ,
  \end{multline}
  where $Z$ is the partition function, $\ket*{r^0}$ are random states in the zero-electron basis, $\ket*{\phi ^0_n}$ are the zero-electron eigenstates and $\ket*{\psi_j}$ are the Lanczos vectors from the one-electron subspace with the corresponding energy $\epsilon_j$. The variational Hilbert space~\cite{bonca99,ku2002} is used together with twisted boundary conditions~\cite{shastry_1990,poilblanc_1991,bonca_03} to limit finite-size effects.
  \subsection{Single-site spectral function and  emission spectrum}
  \label{subsec:ssAp}
  In order to gain a better understanding of the emission spectrum at low temperatures, we derive an analytical expression for the single-site system. This is done for the spectral function in Ref.~\cite{ciuchi97} and a detailed derivation is presented in Ref.~\cite{bonca2019}. This section is a simple extension of that work but is included for self consistency and to guide the discussion.
  The single-site Hamiltonian is
   \begin{equation}
  \label{eq:def_ss_ham}
  \hat H_{\rm s}=\gamma \hat n(\hat b^{\dagger}+\hat b)+\omega_0 \hat b^{\dagger} \hat b.
  \end{equation}
  The solution for the case of a single electron is given by the well-known coherent states
  \begin{equation}
  \label{eq:def_ss_gs}
  \ket{0_{s}}=e^{-\tilde{g}^2/2}\sum\limits_{m=0}^{\infty}\frac{(-\tilde{g})^m}{\sqrt{m!}}\ket{m},
  \end{equation}
  where $\tilde{g}=\gamma/\omega_0$ and $\ket{m}$ are bare phonon modes. The ground state $  \ket{0}_{s}$ has the energy $E_{0,s}=-\omega_0 \tilde{g}^2$. The excited states $  \ket{m}_{s}$ have an energy $E_{m,s}=-\omega_0 \tilde{g}^2+m \omega_0$ and are given by 
    \begin{equation}
   \label{eq:def_ss_gs_energy}
  \ket{m_{s}}=\frac{(b^{\dagger}+\tilde{g})^m}{\sqrt{m!}}\ket{0_{s}}.
  \end{equation}
The single-site spectral function $A_{s}( \omega)$, derived in Ref.~\cite{bonca2019}, is   \begin{equation}
     \label{eq:def_A_ss}
  A_{s}(\omega)=\frac{1}{Z}\sum\limits_{n,m=0}^{\infty}e^{-\beta \omega_0  n} \abs{\ip{m_{s}}{n}}^2 \delta (\omega+E_n -E_{m,s}),
  \end{equation} where $\ket{n}$ are the bare phonon modes and $E_n$ the corresponding energies. We show $ A_{s}(\omega)$ in Fig.~\ref{fig:singlepec}(a) for $\lambda=1$.   

The electron emission spectrum is
    \begin{equation}
   \label{eq:def_Aplus_ss}
  A^+_{s}(\omega)=\frac{1}{Z}\sum\limits_{n,m=0}^{\infty}e^{-\beta E_{m,s}} \abs{\ip{n}{m_{s}}}^2 \delta (\omega+E_n -E_{m,s}).
\end{equation}
The overlap between the coherent state and the normal mode is given by
 \begin{equation}
   \label{eq:def_mn_overlap}
\ip{n}{m_{s}}=e^{-\tilde{g}^2/2}\sum\limits_{l=0}^{\min\{m,n\}}(-1)^{n-l}\tilde{g}^{n+m-2l} \frac{\sqrt{m!n!}}{l!(m-l)!(n-l)!}.
  \end{equation}
  If we send $T/\omega_0 \rightarrow 0$, only the $m=0$ term contributes in Eq.~\eqref{eq:def_Aplus_ss}. This gives 
  \begin{equation}
     \label{eq:def_mn_overlap_T0}
\ip{n}{0_{s}}=e^{-\tilde{g}^2/2} (-1)^{n}\tilde{g}^{n} \frac{\sqrt{n!}}{n!}.
  \end{equation}
  The emission spectrum then takes the form
  \begin{equation}
     \label{eq:def_Aplus_ss_T0}
  A_s^+(\omega)=\sum\limits_{n=0}^{\infty} e^{-\tilde{g}^2} \tilde{g}^{2n} \frac{1}{n!} \delta (\omega+n\omega_0 +\omega_0 \tilde{g}^2),
  \end{equation}
  which has a polaron peak for $n=0$ at $\omega_{\rm pol} / \omega_0 =-\omega_0 \tilde{g}^2$.  The spectrum further has peaks at negative $\omega$, which are separated by $\omega_0$. It is also clear that at larger temperatures, peaks at $\omega>-\omega_0 \tilde{g}^2$ will appear. 

In Fig~\ref{fig:singlepec}(b), we show the single-site emission spectrum $A^+_{s}(\omega)$. There, the peaks at $\omega < \omega_0\tilde{g} $ are visible. One can also observe the peaks at $\omega > \omega_0\tilde{g}$ appearing for larger temperatures.
    \subsection{Single-site  phonon spectral function}
  \label{subsec:ssD}
\begin{figure}[!]
\includegraphics[width=0.99\columnwidth]{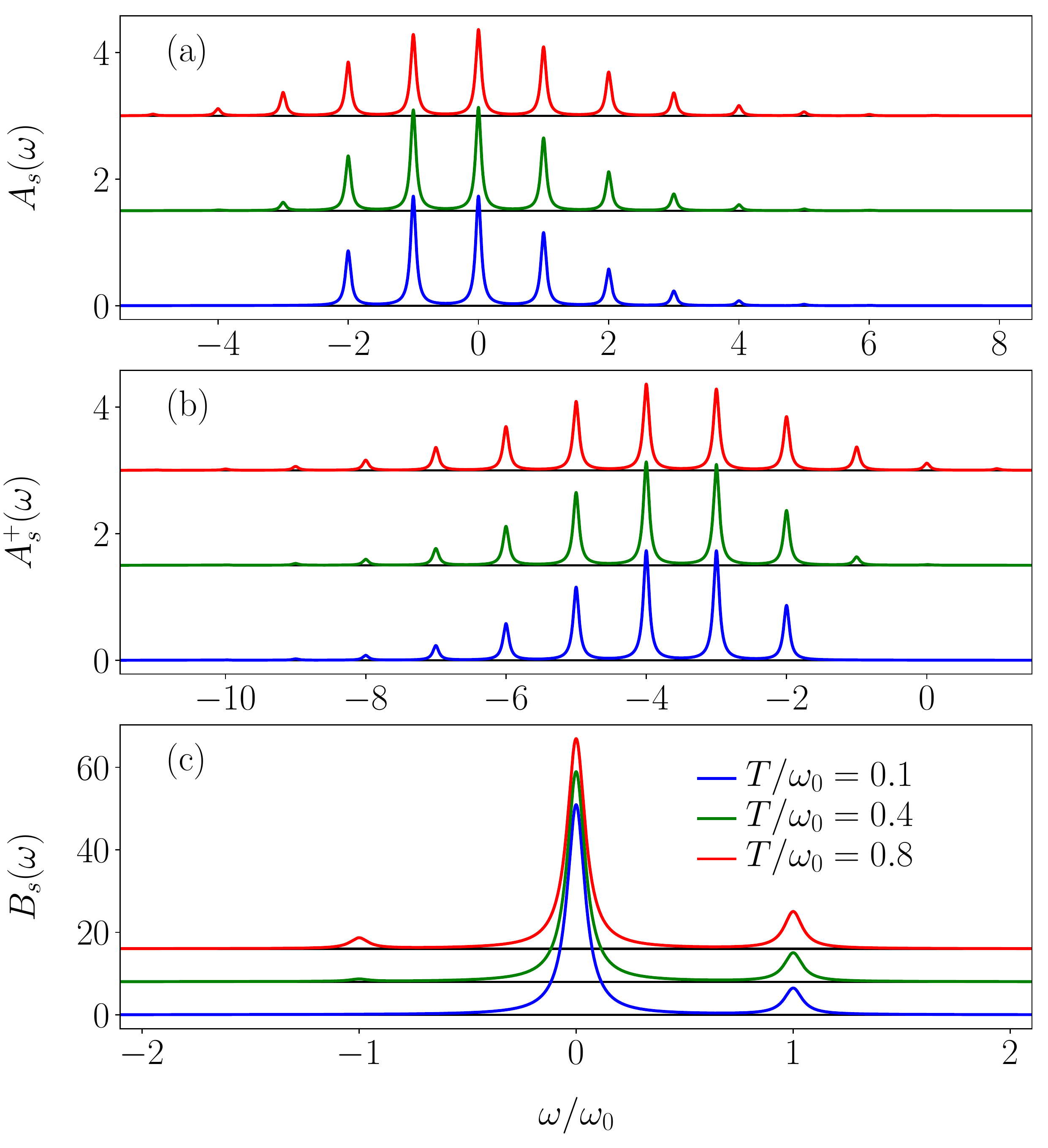}
\caption{(a) Single-site spectral function $A_{s}(\omega)$ from Eq.~\eqref{eq:def_A_ss}.
(b) Single-site emission spectrum  $A_{s}^+(\omega)$ from Eq.~\eqref{eq:def_Aplus_ss}. 
(c) Single-site phonon spectral function $B_{s}^+(\omega)$ from Eq.~\eqref{eq:def_D_ss}. We set $\gamma/\omega_0=\sqrt{2}$ and use a Lorentzian for the delta function with half-width at half-maximum (HWHM) $\eta=0.05$.
}
\label{fig:singlepec}
\end{figure}

  The single-site phonon spectral function is
    \begin{multline}
   \label{eq:def_D_ss}
  B_{s}(\omega)=\frac{1}{Z}\sum\limits_{n,m=0}^{\infty}e^{-\beta E_{m,s}} \abs*{\mel*{n_{s}}{\hat b+\hat b^{\dagger}}{m_{s}}}^2  \\ \times \delta (\omega+E_{m,s}-E_{n,s} ).
\end{multline} 
We use that 
\begin{equation}
     \label{eq:def_appb_on_0ss}
 \hat b \ket{0_{s}}=-\tilde{g}\ket{0_{s}} ,
  \end{equation}
  and 
  \begin{equation}
     \label{eq:def_appb_on_ mss}
 \hat b \ket{m_{s}}=\sqrt{m}\ket{(m-1)_{s}}-\tilde{g}\ket{m_{s}},
  \end{equation}
  to obtain 
  \begin{multline}
     \label{eq:def_sqrauedexp}
  \mel*{n_{s}}{\hat b+\hat b^{\dagger}}{m_{s}}=   \\ \sqrt{n}\bra{(n-1)_{s}}\ket{m_{s}}+\sqrt{m}\bra{n_{s}}\ket{(m-1)_{s}}-2\tilde{g}\bra{n_{s}}\ket{m_{s}}.
  \end{multline}
  At $T/ \omega_0=0$, the spectral function becomes
      \begin{equation}
   \label{eq:def_D_ss_gs}
  B_{s}(\omega)=4\tilde{g}^2\delta(\omega)+\delta(\omega -\omega_0 ).
\end{equation} 
 $B_{s}(\omega)$ is shown for different $T/ \omega_0$ in Fig.~\ref{fig:singlepec}(c). We observe two peaks separated by $\omega_0 $ at the lowest temperatures as predicted in Eq.~\eqref{eq:def_D_ss_gs}. This is the polaron peak and the free-phonon peak. When the temperature is increased, a smaller free-phonon peak starts to appear at $-\omega_0$.
 
\section{DMRG Results for Thermodynamics}
\label{sec:TD}
In this section, we show the results for the thermodynamic quantities introduced in Sec.~\ref{subsec:td}. Thermodynamics for electron-phonon models has already been studied with Monte Carlo methods, see, e.g., Refs.~\cite{scalettar_89,levine_91,weber_18,wang_18} for results for two-dimensional lattices.  As explained in Sec.~\ref{subsec:mps}, the DMRG  purification method starts at $T=\infty$. Finite temperatures are then obtained by imaginary-time evolution. For the Holstein-polaron model with a local phonon cutoff $M$, we have $\expval*{\hat H}_{T=\infty}/(\omega_0 L )=M/2$, such that depending on the phonon number truncation, the imaginary-time evolution will start at a different energy. Additionally, starting points with a finite $M$  are artificial since they do not represent the true $T=\infty$ limit of the system. For this reason, we first want to investigate whether there is a range of temperatures where we can produce states with expectation values that are independent of $M$ for the polaron in the crossover regime $\lambda =1$. Since the results become $M$-dependent and unphysical for large $T/\omega_0$, we choose to focus on $0.1 \leq T/\omega_0 \lesssim 0.4$. Secondly, we want to investigate how the optimal local basis is affected by the imaginary-time evolution.

\begin{figure}[!]
\includegraphics[width=0.99\columnwidth]{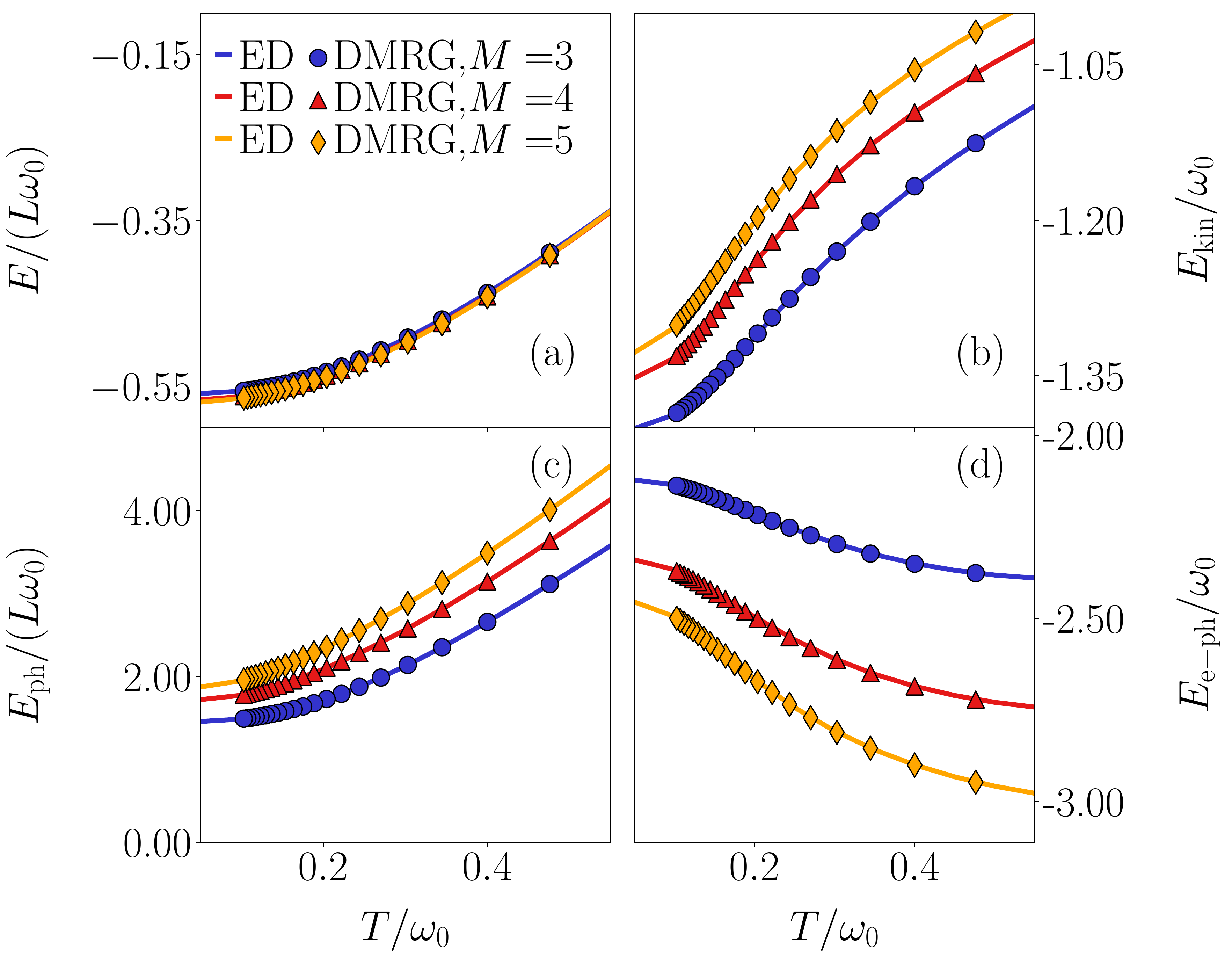}
\caption{Expectation values for different observables for $\gamma/\omega_0=\sqrt{2}, t_0 / \omega_0=1, L=5$ and $M=3,4,5$. The solid lines are obtained with ED and the symbols with DMRG. We show (a) the total energy, (b) the electron kinetic energy, (c) the phonon energy, and (d) the coupling energy. Many points overlap almost completely, and therefore, not all lines and points are visible. For clarity, we only show every fourth point of the DMRG data.}
\label{fig:EDcomp}
\end{figure}

\begin{figure}[!]
\includegraphics[width=0.99\columnwidth]{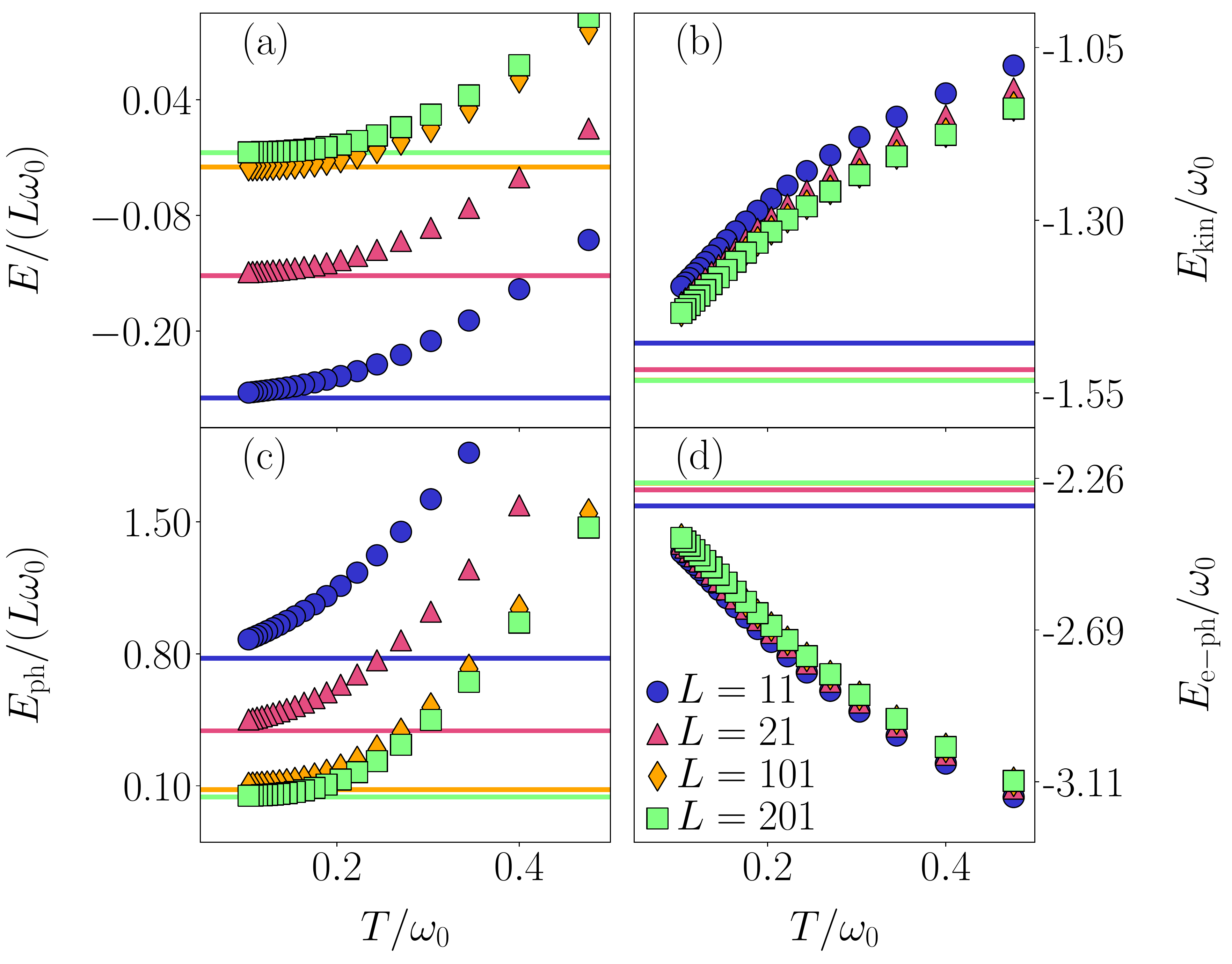}
\caption{Expectation values for different observables for $\gamma/\omega_0=\sqrt{2}, t_0 / \omega_0=1,M=20$ and different $L$. We show (a) the total energy, (b) the electron kinetic energy, (c) the phonon energy, and (d) the coupling energy. The solid lines show the ground-state values calculated with ground-state DMRG. They sometimes overlap, and therefore, some lines are not always visible. For clarity, we only show every fourth data point.}
\label{fig:Lcomp}
\end{figure}

We first verify that the purification method reproduces values calculated with ED.
In Fig.~\ref{fig:EDcomp}, we compare the results to ED for $L= 5$ and different $M$. We show the four observables  defined in Sec.~\ref{subsec:td} as a function of temperature $T/\omega_0$. One sees that the finite-temperature DMRG method (symbols) reproduces the ED (solid lines) for the corresponding $M$. The clear dependence of the observables on $M$ suggests that the local Hilbert space is not chosen large enough to yield the correct low-temperature physics for these parameters. Most importantly, the DMRG method reproduces the ED results for the accessible system sizes.

We proceed by comparing the expectation values for different system sizes $L$. The results are shown in Fig.~\ref{fig:Lcomp}. Notice that the ground-state energy is intensive in the single-electron problem. Therefore, $E/(\omega_0 L) $ should approach zero in the thermodynamic limit. This can be observed in Fig.~\ref{fig:Lcomp}(a). The figure also serves as a consistency check by showing that the imaginary-time evolution approaches  the ground-state energy calculated with ground-state DMRG~\cite{white92,schollwock2005density,schollwock2011density} (solid lines). Both the total energy $E$ and the phonon energy $E_{\rm ph}$ are extensive at finite temperature, and therefore, we divide both of these expectation values by the system size $L$ to get a quantity that only depends on temperature for sufficiently large $L$. The observables $E_{\rm{kin}}$ and $E_{\rm e-ph}$ [Figs.~\ref{fig:Lcomp}(b) and (d)] are automatically intensive since there is only one electron in the system. Figure~\ref{fig:Lcomp} therefore  illustrates that the purification method gives access to thermodynamic quantities in systems with very large local Hilbert spaces.   

\begin{figure}[!]
\includegraphics[width=0.99\columnwidth]{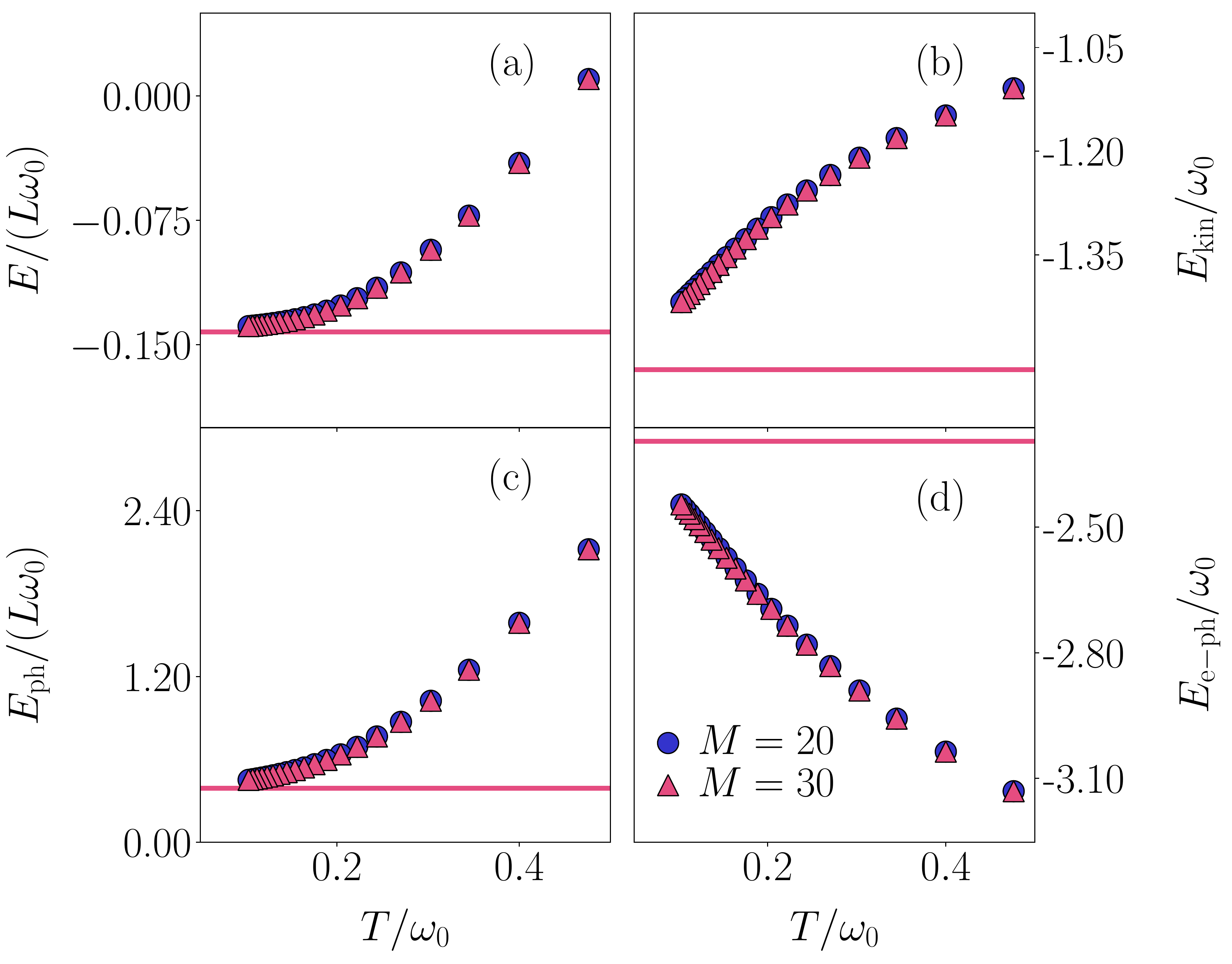}
\caption{Expectation values for different observables for $\gamma/\omega_0=\sqrt{2}, t_0 / \omega_0=1, L=21$ and $M=20,30$. We show (a) the total energy, (b) the electron kinetic energy, (c) the phonon energy,  and (d) the coupling energy. The solid lines show the ground-state values calculated with ground-state DMRG for $M=20$. The points lie on top of each other such that the $M=20$ data is not always clearly visible. For clarity, we only show every fourth point of the data.}
\label{fig:Mconv}
\end{figure}

We next demonstrate that the DMRG method can access values of $M$ large enough to obtain cutoff-independent results in the low-temperature regime.
In Fig.~\ref{fig:Mconv}, we show the same observables as in Fig.~\ref{fig:EDcomp} with $L=21$ and $M=20,30$ calculated with DMRG. We find that even though the two initial states start at two completely different energies $\expval*{\hat H}_{T=\infty}/(\omega_0 L)$, they still converge to the same expectation value up to an accuracy of $O(10^{-5})$ below $T/\omega_0 \lesssim 0.5$. We thus conclude that we correctly reproduce results for the real phonon limit $M \rightarrow \infty$ below a certain temperature if $M$ is chosen large enough. Therefore, the method gives access to thermodynamics at temperatures for system sizes and phonon numbers unavailable to ED and regular Lanczos  methods. For the rest of this paper, we choose $M=20$. 

\begin{figure}[!]
\includegraphics[width=0.99\columnwidth]{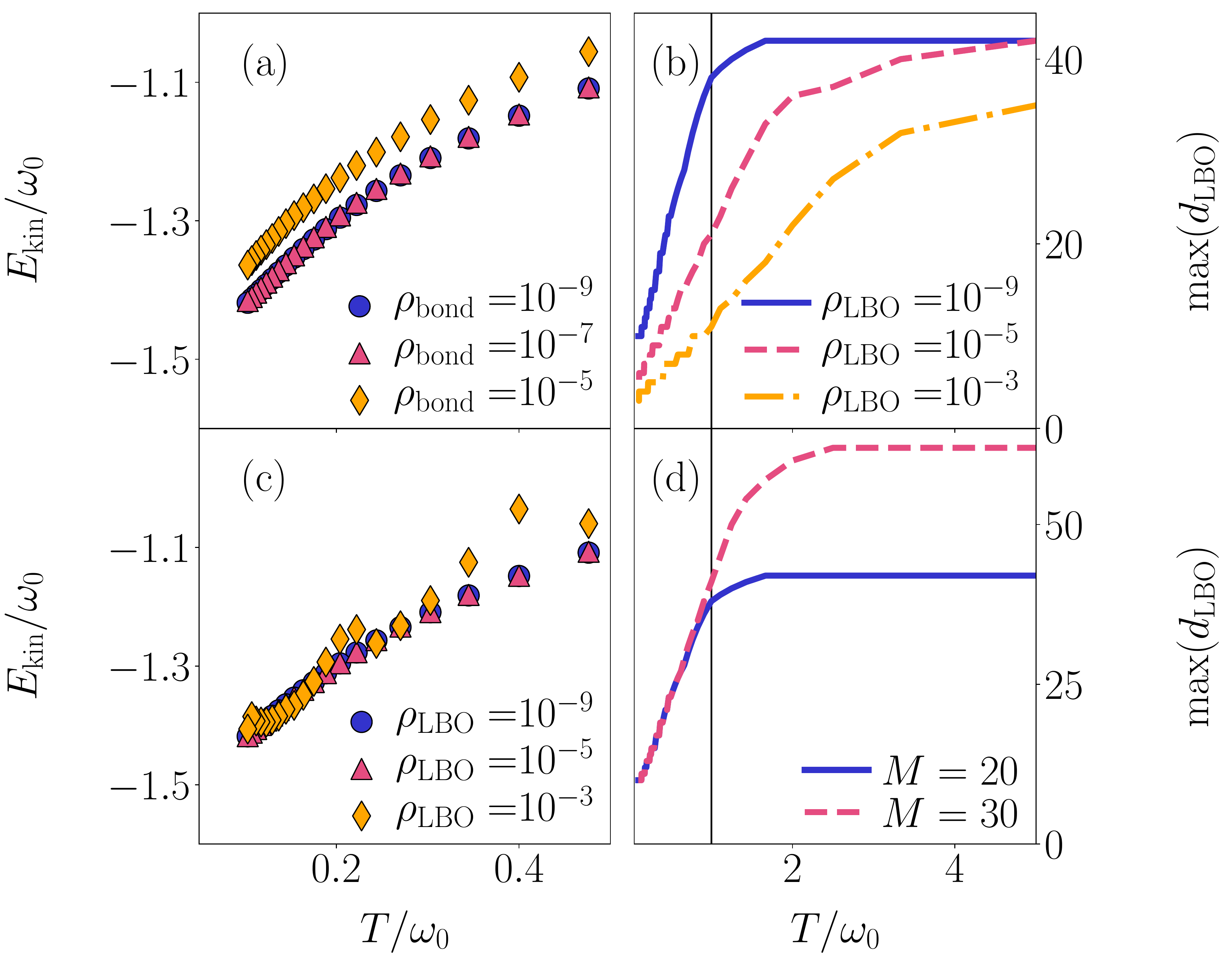}
\caption{Electron kinetic energy from  Eq.~\eqref{eq:def_Hkin} for different $\rho_{\rm bond}$ (a) and  $\rho_{\rm LBO}$ (c). We show results for $\gamma/\omega_0=\sqrt{2}, t_0 / \omega_0=1, L=21$ and $M=20$. In (a), we set $\rho_{\rm LBO}=10^{-9}$ and in (c), we set $\rho_{\rm bond}=10^{-9}$. (b) The maximum local optimal dimension $\max(d_{\rm LBO})$ of the system for the same parameters as (a). (d) $\max(d_{\rm LBO})$ for the same parameters as in (a), but with $\rho_{\rm bond}=10^{-9}$, $M=20$ and $M=30$. The black solid lines in (b) and (d) show $T/\omega_0=1.0$. For clarity, we only show every fourth point of the data in (a) and (c).}
\label{fig:DMRGconv}
\end{figure}

To demonstrate that the imaginary-time evolution results are converged in the low-temperature limit, we vary $\rho_{\rm bond}$ and $\rho_{\rm LBO}$. As explained in Sec.~\ref{sec:method}, the truncation of the bond dimension is controlled by $\rho_{\rm bond}$  whereas $\rho_{\rm LBO}$ controls the truncation of the optimal  local basis of the MPS. In Figs.~\ref{fig:DMRGconv}(a) and (c), we illustrate how $E_{\rm kin}$ is affected by changes in $\rho_{\rm bond}$  and $\rho_{\rm LBO}$. The change is significant if one of the discarded weights is chosen too large. If $\rho_{\rm bond}$ is too large, the expectation values lie above the converged value. In the other case, for $\rho_{\rm LBO}$ too large, we start to get fluctuating expectation values. We do this test for all terms in the Hamiltonian in Eq.~\eqref{eq:def_HolHam} and find that they are all converged for $\rho_{\rm bond}=10^{-7}$ and $\rho_{\rm LBO}=10^{-5}$ to an accuracy of $O(10^{-3})$. However, exactly how they behave for a too small cutoff is observable- and system-size dependent. Since the expectation values already have converged for both $\rho_{\rm bond}$ and $\rho_{\rm LBO}=10^{-7}$ [red triangles in Figs.~\ref{fig:DMRGconv}(a) and (c)], the $10^{-9}$ markers (blue circles) are barely visible. 

In Figs.~\ref{fig:DMRGconv}(b) and (d), we analyze the maximum dimension $\max(d_{\rm LBO})$ of the optimal basis of the physical Hilbert space. One clearly sees that this becomes equal to the local bare basis dimension $2(M+1)$ for large $T/\omega_0$. This is expected since all the phonon modes become equally probable at $T/\omega_0=\infty$. However, $\max(d_{\rm LBO})$ starts to decrease rapidly below a certain temperature and the rotation into the optimal basis becomes computationally beneficial for a given truncation error. This means that the eigenvalues of the reduced-density matrix first do not decay at all until a certain temperature is reached. After that, they start decreasing rapidly as a function of $T/\omega_0$. Figure~\ref{fig:DMRGconv}(d) shows that this trend becomes more pronounced for larger $M$. Furthermore, it illustrates that as $M$ is increased, the rotation into the optimal basis becomes beneficial at higher $T/ \omega_0$.  

The accurate evaluation of thermal expectation values also serves as an important test for the spectral function calculations in Sec.~\ref{sec:specfuncres}.
In Appendix~\ref{sec:app2}, we show that the first temperature-dependent moments can be calculated by either integrating the spectral function or by computing thermal expectation values. We verify the accuracy of the spectral function by comparing both methods. For the rest of this work, we set $\rho_{\rm LBO}=   \rho_{\rm bond}=10^{-9}$ during the imaginary-time evolution. 
\section{Spectral functions}
  \label{sec:specfuncres}
    \subsection{Real-time evolution}
  \label{subsubsec:realt}
\begin{figure}[!]
\includegraphics[width=0.99\columnwidth]{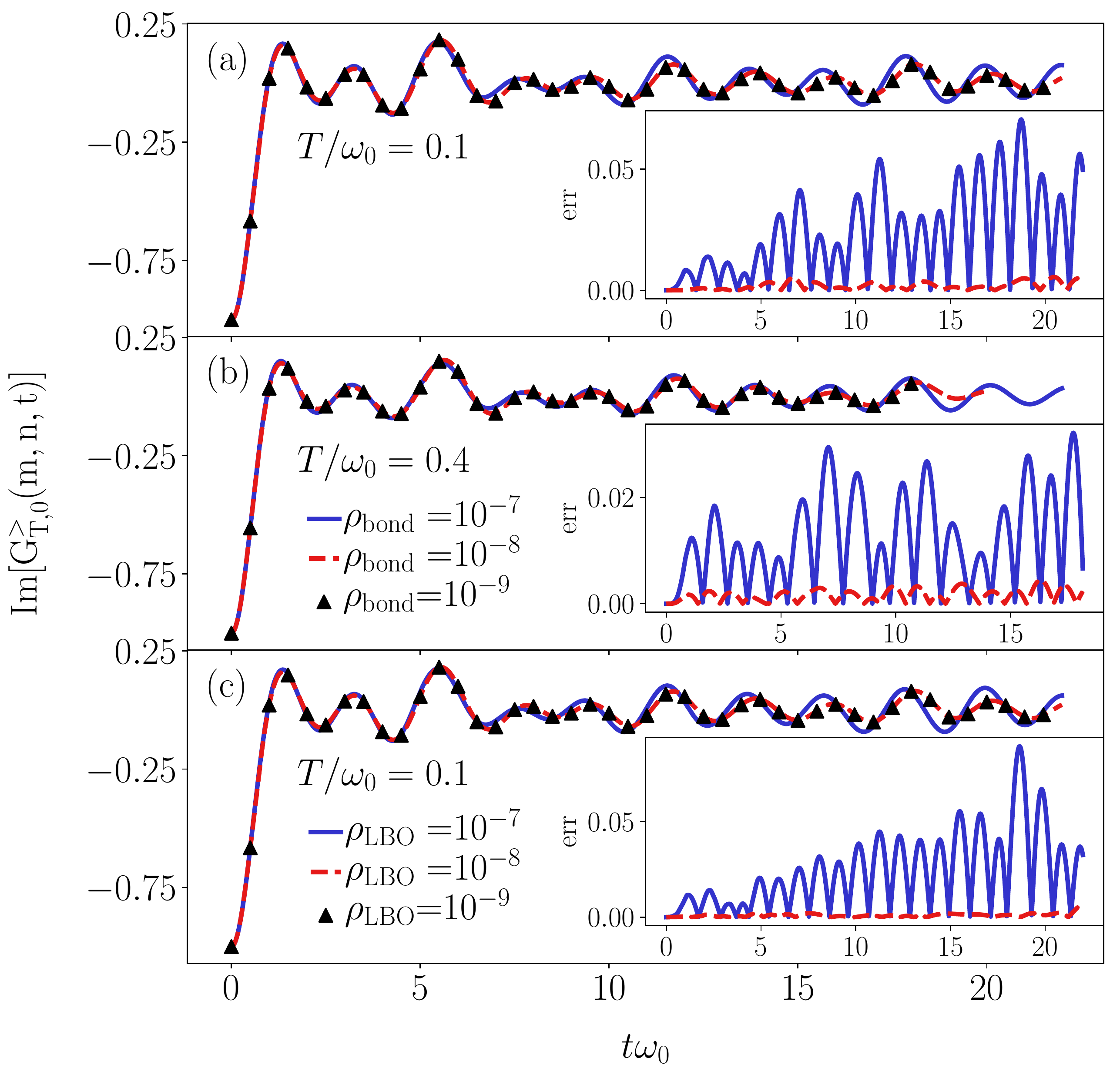}
\caption{Imaginary part of $G_{T,0}^>(m,n, t)$ in Eq.~\eqref{eq:def_corel0}. We set $n=m=11,\gamma/\omega_0=\sqrt{2}, t_0 / \omega_0=1$ and $L=21$.  (a) $T/\omega_0=0.1$  and  fixed $ \rho_{\rm LBO}=10^{-8} $ for different $\rho_{\rm bond}$. (b) $T/\omega_0=0.4$  and  fixed $ \rho_{\rm LBO}=10^{-8} $ for different $\rho_{\rm bond} $. (c) $T/\omega_0=0.1$  and  fixed $ \rho_{\rm bond}=10^{-8} $ for different $\rho_{\rm LBO} $. For the symbols, we only show every 50th point for clarity. The insets show the absolute error $\rm{err}$ [see Eq.~\eqref{eq:def_err_realtime}], defined as the difference between the data with the smallest truncation error $10^{-9}$ and either $10^{-8}$ (red) or $10^{-7}$ (blue).}
\label{fig:Gconv}
\end{figure}

  We now proceed by calculating dynamical properties of our model. We first check that the real-time evolution converges with respect to $\rho_{\rm bond}$ and $\rho_{\rm LBO}$. This is illustrated in Fig.~\ref{fig:Gconv}. There, we show the imaginary part of $G_{T,0}^>(m,n, t)$ from Eq.~\eqref{eq:def_corel0} with $m=n=11$. From Figs.~\ref{fig:Gconv}(a) and (b),
 it becomes apparent that as $\rho_{\rm bond}$ is decreased, the results are indiscernible on the scale of the figure.  A similar behaviour is also seen with respect to $\rho_{\rm LBO}$ [see Fig.~\ref{fig:Gconv}(c)]. In the insets of Fig.~\ref{fig:Gconv}, we show the absolute error 
 \begin{equation}
 \label{eq:def_err_realtime}
 \rm{err}= \abs*{\Im[G_{T,0}^>(m,n, t)]_{\rho_i}-\Im[G_{T,0}^>(m,n, t)]_{\rho_j}},
\end{equation} 
  with $i \neq j$ and $\rho_i$ being set by $\rho_{\rm bond}$ or $\rho_{\rm LBO}$. For the data shown in the inset, we fix $\rho_i=10^{-9}$ and subtract the remaining two datasets. We can also report a large increase in the bond dimension as the temperature is increased (for details, see Appendix \ref{sec:app1}). This is the reason why the time evolution for certain values of $\rho_{\rm bond}$ in Fig.~\ref{fig:Gconv}(b) is stopped earlier. The reachable time for the smallest truncation also determines $t_{\rm max}$, such that the convergence is tested for the whole time interval used for the Fourier transformation. The limitation in accessible times also constrains the energy resolution of  the spectral function.
  
  As explained in Sec.~\ref{sec:method}, we additionally apply linear prediction. The spectral functions tend to oscillate around zero away from the peaks as a result of the finite time interval. By applying linear prediction, the oscillation amplitude goes from order $10^{-1}$ to  $10^{-5}$ without changing the peak position or height in the spectrum. However, the exact decrease of the amplitude is spectral-function and temperature dependent. Due to the oscillations, we always show the absolute values of the spectral functions in the normalized log-scaled plots. The real-time evolution for all the following spectral functions is done with $\rho_{\rm bond}=\rho_{\rm LBO}=10^{-8}$.
  
  To test the accuracy of the method, we also derive the first temperature-dependent moment of the spectral functions and compare the thermal expectation values to our numerical data. The results in Appendix~\ref{sec:app2} show good agreement. We further want to emphasize that in the zero-electron sector, it is trivial to obtain the finite-temperature initial state for the real-time evolution since it only contains non-interacting local harmonic oscillators. For this reason, we compare results using both the trivially obtained thermal states and those obtained with the imaginary-time evolution algorithm to verify its correctness.
   \subsection{Electron spectral function and comparison to the FTLM}
   \label{subsubsec:espec}
        \begin{figure}[t]
\includegraphics[width=0.99\columnwidth]{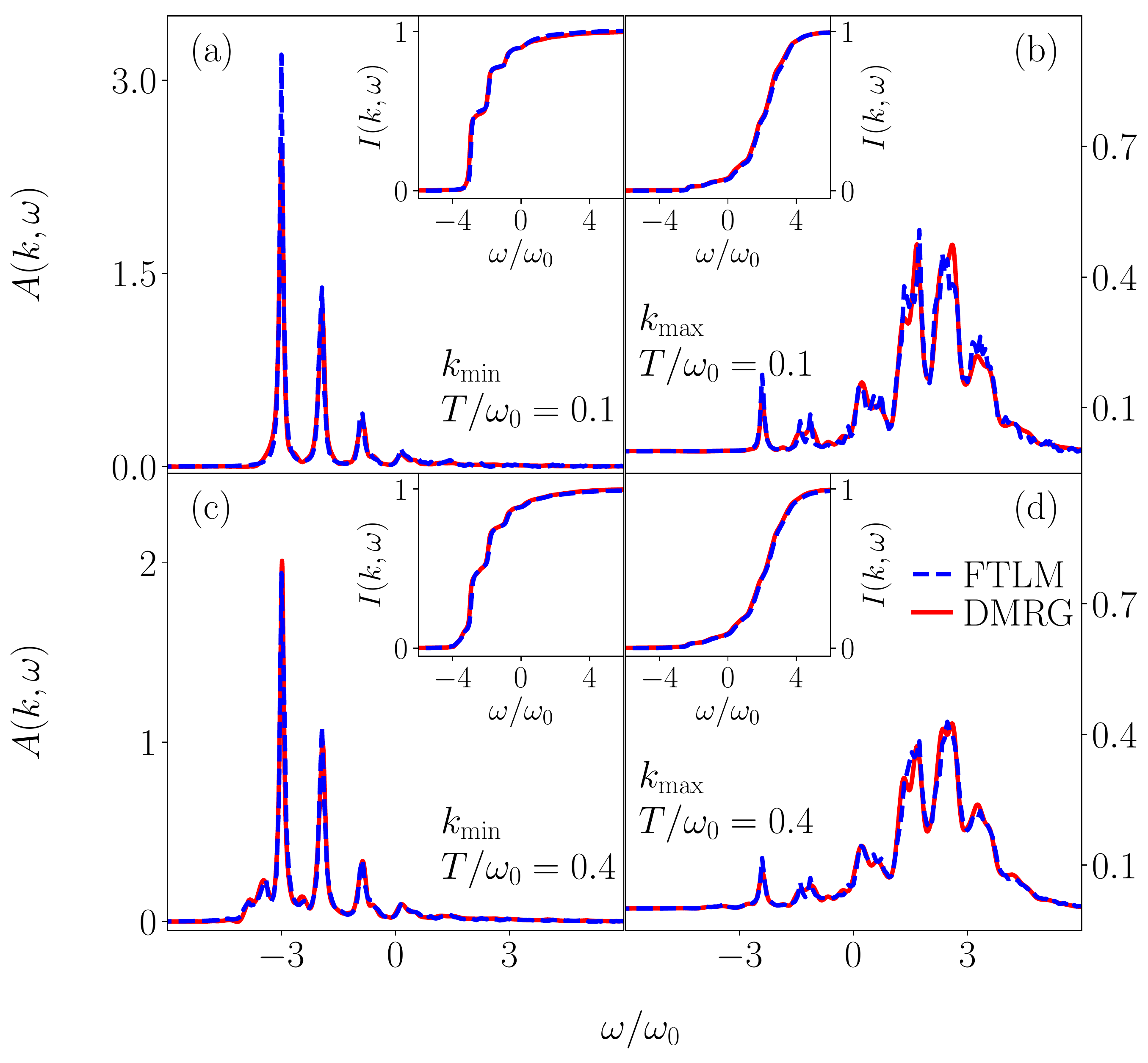}
\caption{Electron spectral function  $A(k, \omega)$ from Eq.~\eqref{eq:def_ferspec} for $\gamma/\omega_0=\sqrt{2}, t_0 / \omega_0=1, L=21, M=20,\eta=0.05$ and $t_{\rm max}\omega_0=18.0$. We show $T/\omega_0=0.1$ in (a) and (b) and $T/\omega_0=0.4$ in (c) and (d). For the DMRG data (red line), we show $k_{\min }=\pi/(L+1)$ and $k_{\max }=\pi L/(L+1) $.  For the FTLM data (blue dashed line), $L=12$ and we show $k_{\min }=0$ in (a) and (c) and $k_{\max }=\pi $ in (b) and (d). The insets show $I(k, \omega)$ defined in Eq.~\eqref{eq:def_I}.}
\label{fig:G0spec}
\end{figure} 

 In Fig.~\ref{fig:G0spec}, we show the electron spectral function $A( k,\omega)$ and compare it to results obtained with FTLM. We show the results for $T/ \omega_0 = 0.1$ in Figs.~\ref{fig:G0spec}(a) and (b) and $T/ \omega_0 = 0.4$ in Figs.~\ref{fig:G0spec}(c) and (d). Our method can resolve the same peak positions as the FTLM. One can identify the polaron peak at $\omega_{\rm pol} / \omega_0\approx-3.0$ and the peaks corresponding to the polaron with additional phonons separated by $n \omega_0$ in the incoherent part of the spectrum. We also observe a significant decrease of the quasi-particle weight for $k_{\rm min}$ compared to $k_{\rm max}$.  This has already been reported in Ref.~\cite{bonca2019} and is consistent with other ground-state approaches~\cite{fehske_2000,hohenadler_03,lau_07,goodvin06,berciu_goodvin_07}. 

 In the inset, we show 
 \begin{equation}
 \label{eq:def_I}
 I(k, \omega)=\int_{-\infty}^{\omega}d \omega^{\prime}A(k, \omega^{\prime}).
 \end{equation} 
There are only small differences between the FTLM and the DMRG data. The amplitude of the polaron peak exhibits small temperature-dependent differences between the two methods. For larger values of $k$,  we also observe some different weight distribution in the incoherent part of the spectrum. The results for $I(k, \omega) $ still almost completely overlap. We want to emphasize that we show results for two different $k$ values for the methods due to the difference in boundary conditions. For the DMRG method, we choose $k_{\rm min}=\pi /(L+1), k_{\rm max}=\pi L /(L+1)$ and for the FTLM method, we select $k_{\rm min}=0, k_{\rm max}=\pi$. We conclude that despite these differences, the DMRG and FTLM method show a very good quantitative agreement.

\begin{figure}[t]
\includegraphics[width=0.99\columnwidth]{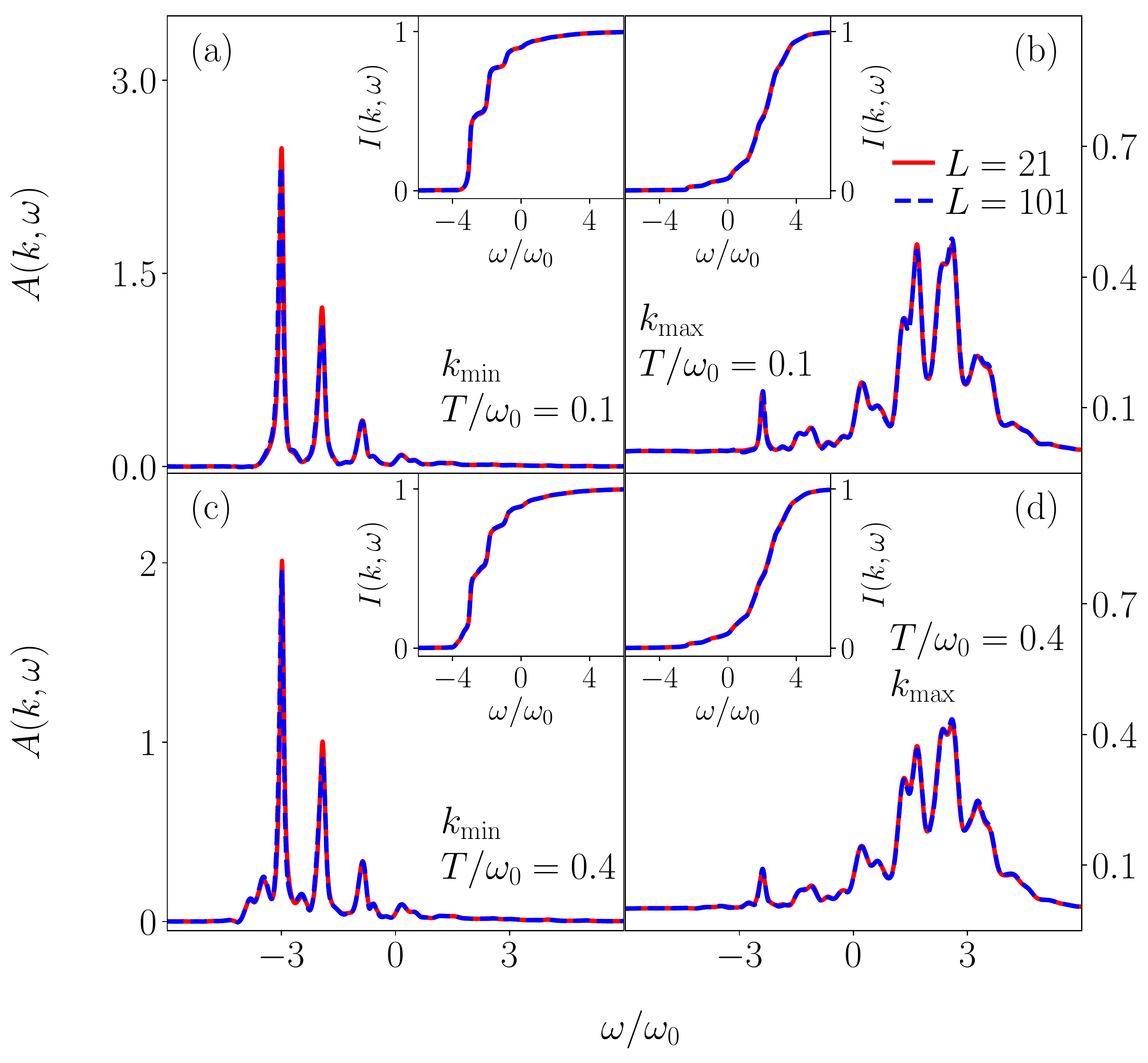}
\caption{Electron spectral function  $A(k, \omega)$ from Eq.~\eqref{eq:def_ferspec} for $\gamma/\omega_0=\sqrt{2}, t_0 / \omega_0=1,M=20,\eta=0.05$ and $t_{\rm max }\omega_0=18.0$. We show $T/\omega_0=0.1$ in (a) and (b) and $T/\omega_0=0.4$ in (c) and (d). For $L=21$, we show $k_{\min }=\pi /(L+1)$ and $k_{\max }=\pi L/(L+1) $. The $L=101$ data (blue dashed lines) is calculated with the simplified Fourier transform (see the text for details) and $k_{\min }=0$ and $k_{\max }=\pi $. The insets show $I(k, \omega)$ defined in Eq.~\eqref{eq:def_I}.}
\label{fig:G0specdL}
\end{figure}

 Alternatively to computing the complete correlation function, one can calculate $G_{T,0}^>(m,n, t)$ for a fixed $n=L/2$ and $m \leq n$. This gives access  to much larger system sizes, as illustrated in Fig.~\ref{fig:G0specdL}. There, we calculate the spectral function for $L=101$. We first fix $n=51$ and set $G_{T,0}^>(n+m,n, t)=G_{T,0}^>(n-m,n, t)$. We then compute the Fourier transform into $k$-space as $G_{T,0}^>(k, t)=\frac{1}{L}\sum\limits_{m=1}^L e^{i(m-n)k}G_{T,0}^>(m,n, t)$. Here, we use  periodic boundary-condition quasi-momenta with $k=2 \pi m/L$ and $-\frac{L}{2} \leq m \leq \frac{L}{2}$. This is often done (e.g., in Refs.~\cite{feiguin2010,paeckel_fauseweh_19}) and the method works well here since the noninteracting harmonic oscillators are homogeneously distributed in the initial state and there is no electron in the system. In Fig.~\ref{fig:G0specdL}, we show comparison between data produced with periodic-boundary condition momenta ($L=101$) with results for open-boundary momenta ($L=21$). Only small changes in the largest peaks [see Figs.~\ref{fig:G0specdL}(a) and (c)] can be seen even though the $L=101$ data use $k_{\rm min}=0, k_{\rm max}=\pi$ while the $L=21$ data use $k_{\rm min}=\pi /(L+1), k_{\rm max}=\pi L /(L+1)$. This is, however, not the case for the other spectral functions studied in this work since  the one-electron state has an inhomogeneous electron distribution. The previously described approach does, therefore, not fulfil the sum rules in those cases. Moreover, we mention that the calculations with the periodic boundary-condition Fourier transformation is more sensible to the choice of parameters for the linear prediction for our data. For a more quantitative discussion of the error of the methods, see Appendix~\ref{sec:app0}.
 
   In Fig.~\ref{fig:G0dens}, we show  $A(k, \omega)$ as a function of $\omega$ for all $k$. Here, the spectral weight at $\omega<\omega_{\rm pol}$ [see Fig.~\ref{fig:G0spec}(a)] becomes visible at larger $T/ \omega_0$. This has been reported in Ref.~\cite{bonca2019} and corresponds to the electron absorbing a thermal phonon. One can also see that the polaron band structure is shifted downwards and renormalized compared to the free-fermion case which would have its ground-state energy at $\omega /\omega_0=-2$ and a bandwidth of $4t_0$. In all cases, we confirm that the sum rule $\int_{-\infty}^{\infty}d\omega A(k, \omega)=1$ is fulfilled up to $10^{-5}$.  

     \begin{figure}[t]
\includegraphics[width=0.99\columnwidth]{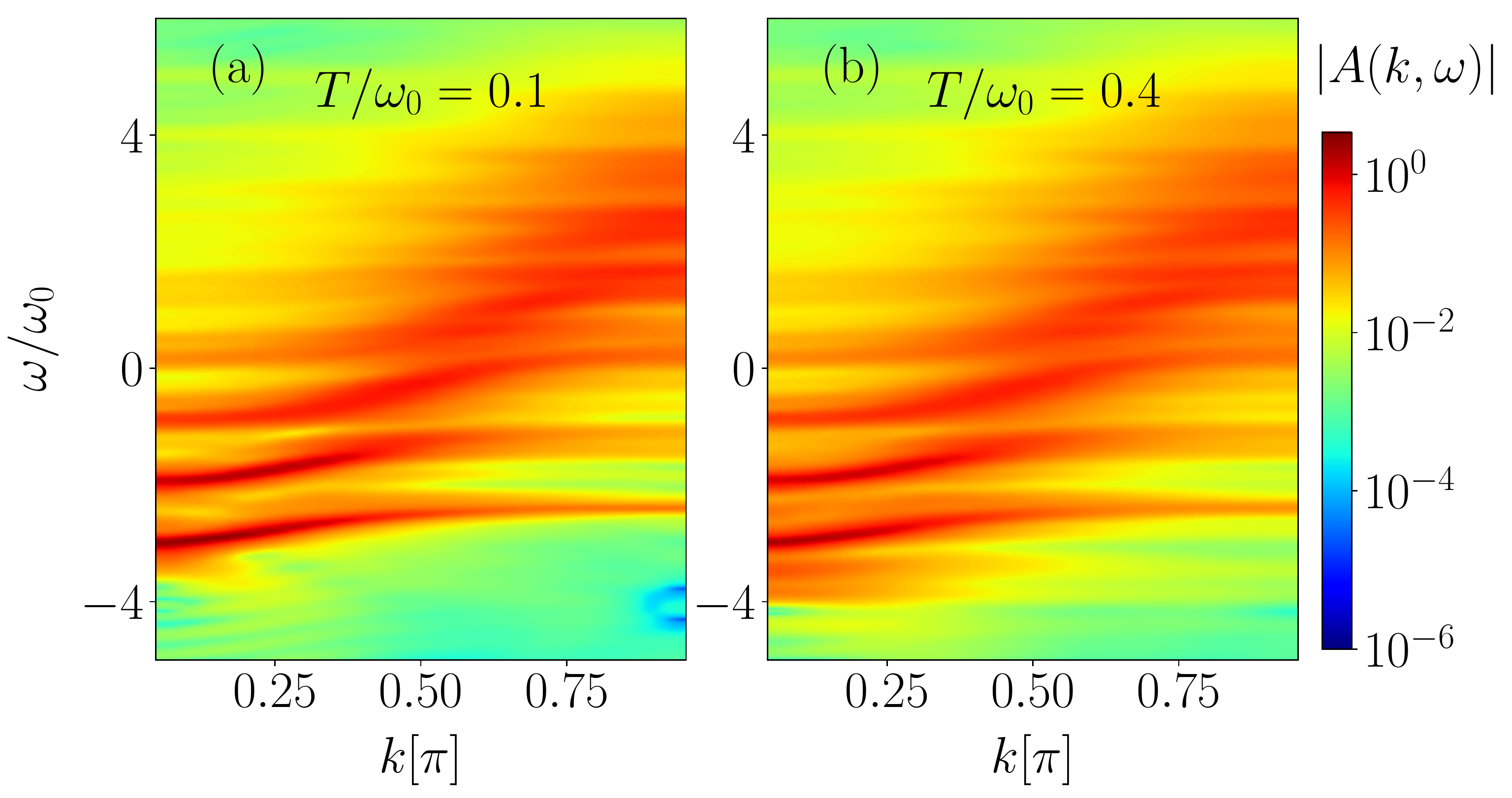}
\caption{Electron spectral function  $A(k, \omega)$ obtained with DMRG. We set $\gamma/\omega_0=\sqrt{2}, t_0 / \omega_0=1, L=21, M=20,\eta=0.05$ and $t_{\rm max}\omega_0=18.0$. (a) $T/ \omega_0=0.1$ and (b)  $T/\omega_0=0.4$. }
\label{fig:G0dens}
\end{figure}
  \subsection{Electron emission spectrum}
  \label{subsubsec:eem}
\begin{figure}[!]
\includegraphics[width=0.99\columnwidth]{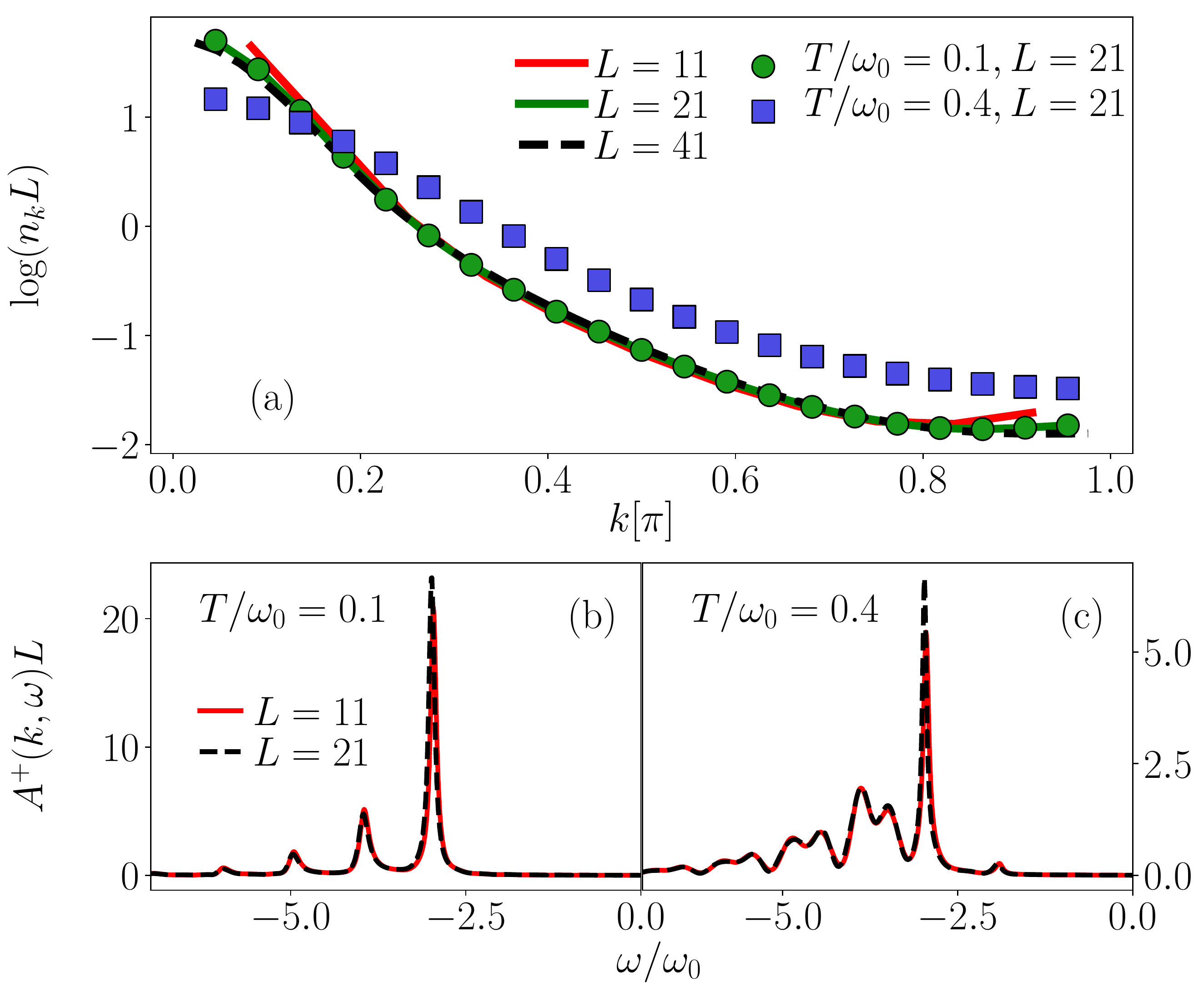}
\caption{(a) Momentum distribution $n_k$ calculated with $\gamma/\omega_0=\sqrt{2}, t_0 / \omega_0=1, M=20,t_{\rm max} \omega_0=18.0, \eta=0.05$ for different system sizes at different temperatures. The symbols show $n_k$ extracted from the Fourier transformed data [see Eq.~\eqref{eq:def_nkFT}] and the solid lines were obtained by calculating the expectation value $\expval{\hat n_k}_T$ at $T/\omega_0 =0.1$. (b) and (c) show the electron emission spectrum $A^+(k, \omega)$ defined in Eq.~\eqref{eq:def_spectild} for the same parameters as in (a) for $T/ \omega_0=0.1$, $T/ \omega_0=0.4$ and $k=\pi/(L+1)$. We show $L=11$ (red solid line) and $L=21$ (black dashed line).  }
\label{fig:AEM}
\end{figure}

\begin{figure}[!]
\includegraphics[width=0.99\columnwidth]{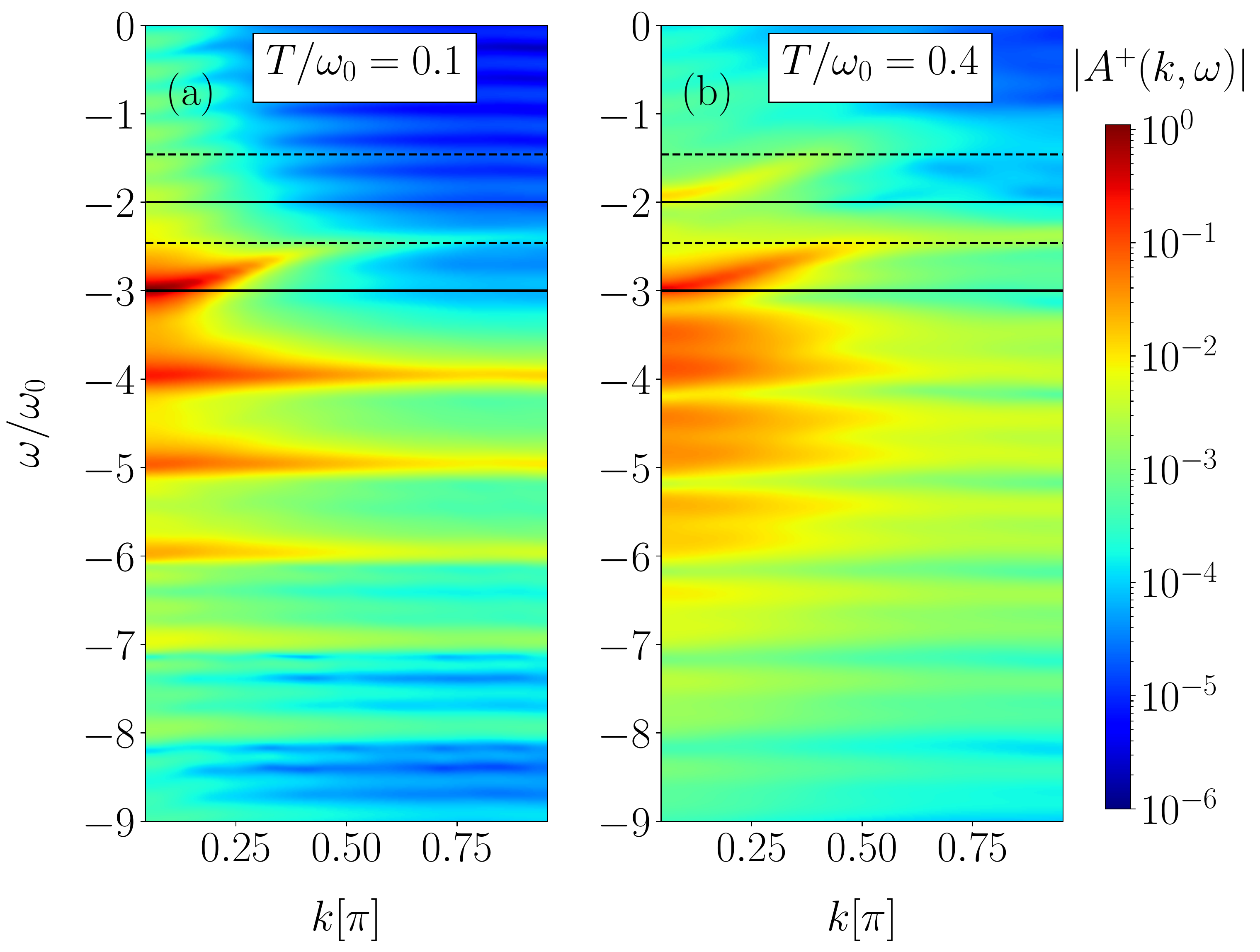}
\caption{Electron emission spectrum  $A^+(k, \omega)$ defined in Eq.~\eqref{eq:def_spectild} obtained with DMRG. The parameters are $\gamma/\omega_0=\sqrt{2}, t_0 / \omega_0=1, L=21, M=20, \eta=0.05$ and $t_{\rm  max}\omega_0=18.0$ for (a)  $T/\omega_0=0.1$ and (b) $T/\omega_0=0.4$. The solid lines show $\omega / \omega_0=-3$ and $\omega / \omega_0=-2$. The dashed lines show $\omega / \omega_0=-3+ 4e^{-\tilde{g}^2}$ and $\omega / \omega_0=-2 +4e^{-\tilde{g}^2}$.}
\label{fig:AEMdens}
\end{figure}

  We next discuss $A^+(k, \omega)$, defined in Eq.~\eqref{eq:def_spectild}. Computationally, this function is the easiest to obtain with our method since the most demanding part of the calculation, namely the real-time evolution, is done without an electron in the physical system. In Fig.~\ref{fig:AEM}, we show $A^{+}(k, \omega)$  for two different temperatures. At low $T/\omega_0$,  Fig.~\ref{fig:AEM}(b) unveils the presence of  several peaks that are separated by $\omega_0$. The peaks can be understood by inspecting the single-site emission spectrum at low temperatures from Eq.~\eqref{eq:def_Aplus_ss_T0}. There, one clearly sees a peak at $-\gamma^{2}/\omega_0$. Furthermore, there are several peaks at negative $\omega$ separated with $\omega_0$. In Fig.~\ref{fig:AEM}(b), we have one main peak at the ground-state energy $\omega_{\rm pol}/\omega_0 \approx -3$. This peak is also robust against an increase in temperature [see Fig.~\ref{fig:AEM}(c)]. The peaks at lower $\omega $, however, acquire more structure at elevated $T/ \omega_0$. We also observe a peak at $\omega_{\rm pol}/\omega_0 +1$ which is completely suppressed at $T/\omega_0 =0.1$. In Fig.~\ref{fig:AEMdens}, the complete function $A^+(k, \omega)$ is plotted as a function of $k$ and $\omega$. At $T/\omega_0 =0.4$ [Fig.~\ref{fig:AEMdens}(b)], we see two clear polaron bands starting at $\omega / \omega_0 =-3 $ and $\omega / \omega_0 =-2 $. Both have a bandwidth of $D\approx e^{-\tilde{g}^2}4\approx 0.54$, which is illustrated by the black dashed and solid lines. The peaks at lower frequencies also seem to shift towards higher frequencies and additional peaks appear to emerge at approximately  $D$ away from the already existing ones at $\omega / \omega_0 <-3$.  
  
  In Fig.~\ref{fig:AEM}(a), we show the electron momentum distribution calculated for different system sizes. This quantity can be calculated directly as the thermal expectation value $\expval*{n_k}_T$ or extracted from the lesser Green's function 
    \begin{equation} \label{eq:def_nkFT}
n_k=\int_{-\infty}^{\infty} d\omega A^+(k, \omega) \, .
\end{equation} 
As a consistency check, we show both. Figure~\ref{fig:AEM}(a) illustrates that some finite-size effects exist for small $L$, but as $L$ is increased, $n_kL$ converges. Note that with increasing $L$, the number of $k$-points also increases, however, $\sum_k n_k =1$. This is, of course, different for the spectral function 
discussed in Sec.~\ref{subsubsec:espec}, where
\begin{equation} \label{eq:def_n_partcon}
 \int_{-\infty}^{\infty} d\omega A(k, \omega) =1\, 
\end{equation} 
for all $k$ and $L$.  We thus conclude that already at the low temperatures studied here, $n_k$ starts to flatten out and the difference in amplitude between the polaron peak and the other peaks decrease.

  \subsection{Phonon spectral function}
  \label{subsubsec:phspec}
    \begin{figure}[t]
\includegraphics[width=0.99\columnwidth]{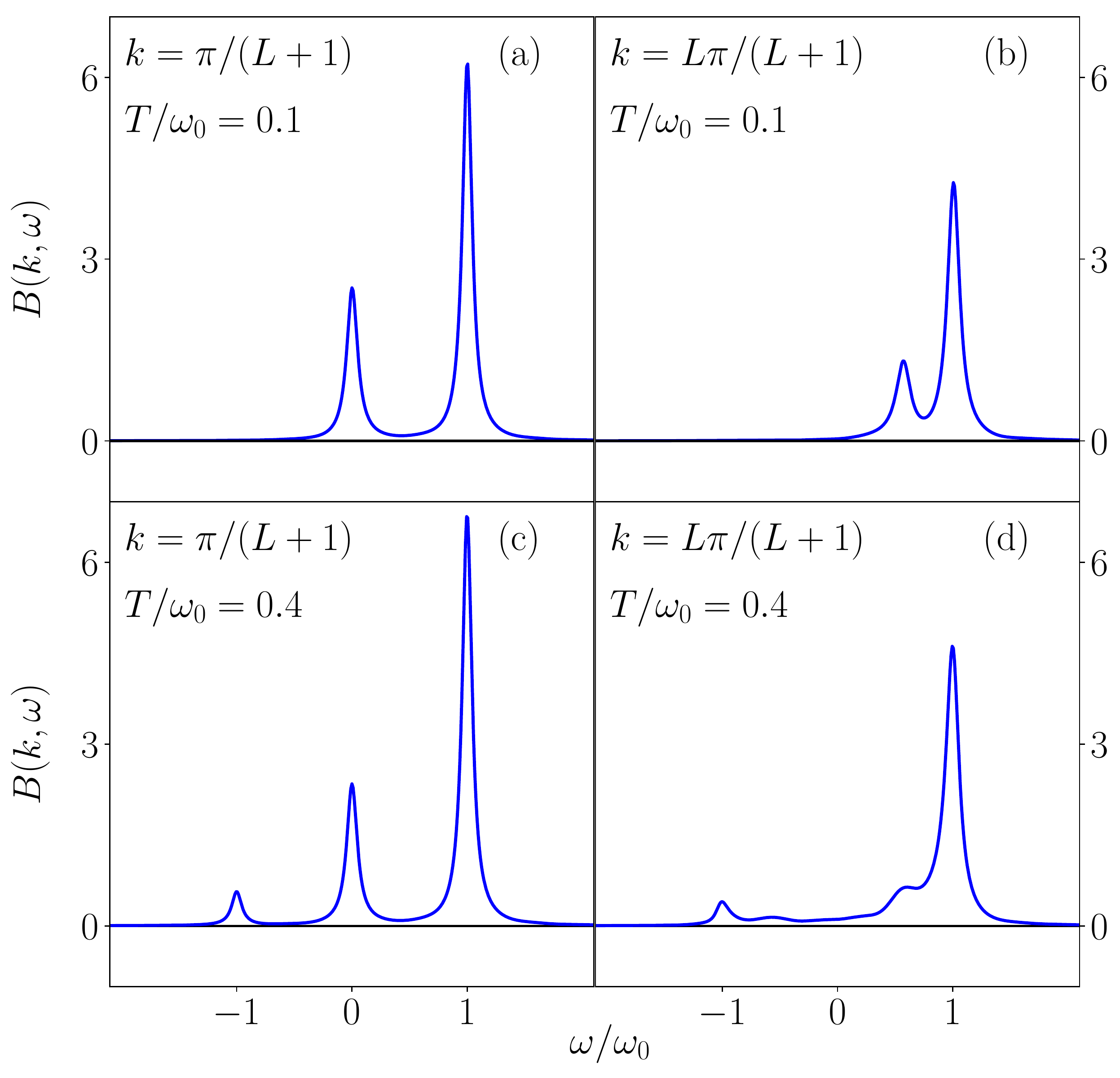}
\caption{Phonon spectral function $B(k, \omega)$ from Eq.~\eqref{eq:def_phospec} obtained with DMRG. The parameters are $\gamma/\omega_0=\sqrt{2}, t_0 / \omega_0=1, L=21,M=20,\eta=0.05 $ and $t_{\rm max}\omega_0=15.8$. (a) and (b) show $T / \omega_0=0.1$ with $k=\pi/(L+1)$ and $k=L\pi/(L+1)$. (c) and (d) show the same $k$-values for $T /\omega_0=0.4$.}
\label{fig:Bspec}
\end{figure}

We now move on to the phonon spectral function. Its ground-state properties have already been studied thoroughly (see,  e.g., Refs.~\cite{loos_2006,vidmar10}). In Ref.~\cite{loos_2006}, Loos \textit{et al.}~used analytic and numerical methods to study this spectral function in a variety of parameter regimes. They found that the dominating features of the phonon spectral function are a free-phonon line and a renormalized band dispersion with additional structure appearing for intermediate electron-phonon coupling. Vidmar \textit{et al.}~\cite{vidmar10} studied the low-energy spectrum and identified several bound and anti-bound states in different parameter regimes.

Here, we are interested in this function at finite temperature and in  Fig.~\ref{fig:Bspec}, we display $B(k, \omega)$ for $T/\omega_0=0.1$ and $0.4$ for different $k$. At $T/\omega_0=0.1$, which is close to the ground state, we clearly recognize two distinct peaks, one at $\omega /\omega_0 =1 $ and another one that gets shifted with $k$. The peak at $\omega /\omega_0 =1 $ originates  from the free phonon, whereas the other peak originates from the phonon being coupled to the electron. When temperature is increased, the phonon spectral function changes dramatically. For $k=L\pi/(L+1)$, the peaks get significantly broader and the  polaron and the free-phonon peaks are almost completely merged. We also see structure appearing at $\omega /\omega_0<0$ which is suppressed at low temperatures, exactly as is the case for the single-site phonon spectral function in Sec.~\ref{subsec:ssD}. 

\begin{figure}[t]
\includegraphics[width=0.99\columnwidth]{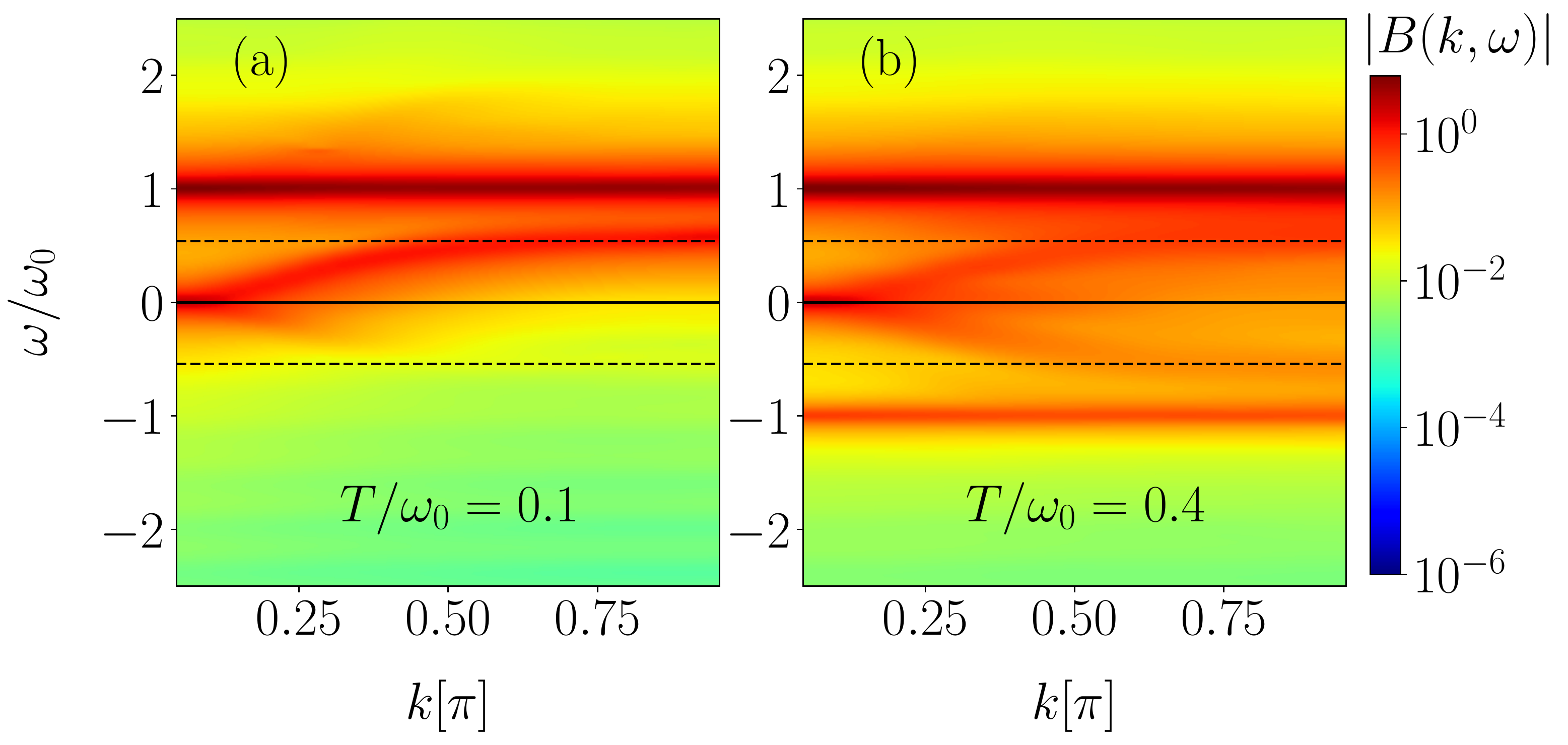}
\caption{Phonon spectral function $B(k, \omega)$ from Eq.~\eqref{eq:def_phospec} obtained with DMRG. The parameters are $\gamma/\omega_0=\sqrt{2}, t_0 / \omega_0=1, L=21,M=20, \eta=0.05$ and $t_{\rm max}\omega_0=15.8$. (a) shows $T / \omega_0=0.1$ and (b) shows $T /\omega_0=0.4$. The solid lines show $\omega / \omega_0=0$ and the dashed lines show $\omega / \omega_0=\pm 4e^{-\tilde{g}^2}$.}
\label{fig:Bspecdense}
\end{figure}

 In Fig.~\ref{fig:Bspecdense}, we show the complete $B(k, \omega)$. Here, the polaron band structure with a width of $D\approx 4e^{-\tilde{g}^2}$ is visible. It also becomes clear that whereas we only observe the appearance of a free-phonon peak at negative frequencies in the single-site case, here, there is a complete reflected polaron band appearing for $T/\omega_0=0.4$ at $\omega/ \omega_0<0$ [see Fig.~\ref{fig:Bspecdense}(b)]. This is similar to what we find for the emission spectrum in Fig.~\ref{fig:AEMdens}.
\section{Summary} \label{sec:summary}
We have generalized the DMRG method combined with purification and local basis optimization to efficiently compute static as well as dynamic properties of the Holstein polaron in the intermediate coupling regime  at finite temperatures. We first showed that the method enabled us to  generate thermal states at a finite temperature by performing imaginary-time evolution. We then computed the electron  spectral function and showed that our results quantitatively agree  with those obtained using the finite-temperature Lanczos method of Ref.~\cite{bonca2019}. We also analyzed the electron emission spectrum and found that the difference between the amplitude of the polaron peak and the other peaks decreased and that $n_k$ flattened out with increasing  temperature. In addition, we observed an additional band appearing at larger $\omega$. Regarding  the phonon spectral function, our work unveils that with increasing temperature, the spectrum broadens at larger momentum accompanied by the emergence of a mirrored image at $\omega <0$.

We propose a number of future applications of the method introduced in this work. A natural extension would be to compare the results presented in this work to similar calculations done with minimally entangled typical thermal state algorithms \cite{white2009,stoudenmire2010,binder_15,Bruognolo_17,goto_19,Agasti_2020}. Another  direction would be to combine the local basis optimization with other time-evolution methods~\cite{haegeman_11,haegeman_16,yang_20,kloss_19,secular_2020}. A further possible area of application is to calculate thermal expectation values combined with quench dynamics~\cite{murakami_werner_15,brockt_dorfner_15,nabeyendu_16,wall_16,shroeder_16,brockt_17,shota_18,mannouch_18,stolpp2020}
 to test the predictions of the eigenstate thermalization hypothesis~\cite{rigol_dunjko_08,rigol_srednicki_12,sorg14,dalessio_kafri_16,kogoj_vidmar_16,deutsch_18,mori_ikeda_18,jansen19}. The proposed method can also be generalized to investigate hetero-junctions containing vibrational degrees of freedom~\cite{galperin_2007,osorio_2008,Koch_2010,zimbovskaya_2011,khedri_18,wang_18,dey_18} and to study the evolution of polaron states in manganites~\cite{westerhauser_2006,sheu14,raiser_17,sotoudeh_17}.
  Further challenging continuations could involve the numerical study of time-dependent spectral functions (see, e.g.,~\cite{paeckel_fauseweh_19,freericks_09}) relevant to time-dependent ARPES experiments~\cite{damascell_03,damascelli_04,eckstein_08,ligges_18} or to compute the optical conductivity at finite temperatures (see, e.g., Refs.~\cite{goodvin11,mendoza-arenas_19,fetherolf_20}).

\section*{Acknowledgment} \label{sec:ack}
We  acknowledge  useful  discussions with A. Feiguin, E. Jeckelmann, C. Karrasch,  S. Manmana, C. Meyer, S. Paeckel and J. Stolpp.
D.J. and F.H.-M. were  funded by the Deutsche Forschungsgemeinschaft (DFG, German Research Foundation) – 217133147 via SFB 1073 (project B09). J.B.   acknowledges the support by the program P1-0044 of the
Slovenian Research Agency, support from  the Centre for Integrated Nanotechnologies, a U.S. Department of Energy, Office of Basic Energy Sciences user facility, and funding from the Stewart Blusson Quantum Matter Institute. 
\\

\appendix
\section{Bond and LBO dimension in real-time evolution} \label{sec:app1}
   \begin{figure}[t]
    \centering
\includegraphics[width=0.99\columnwidth]{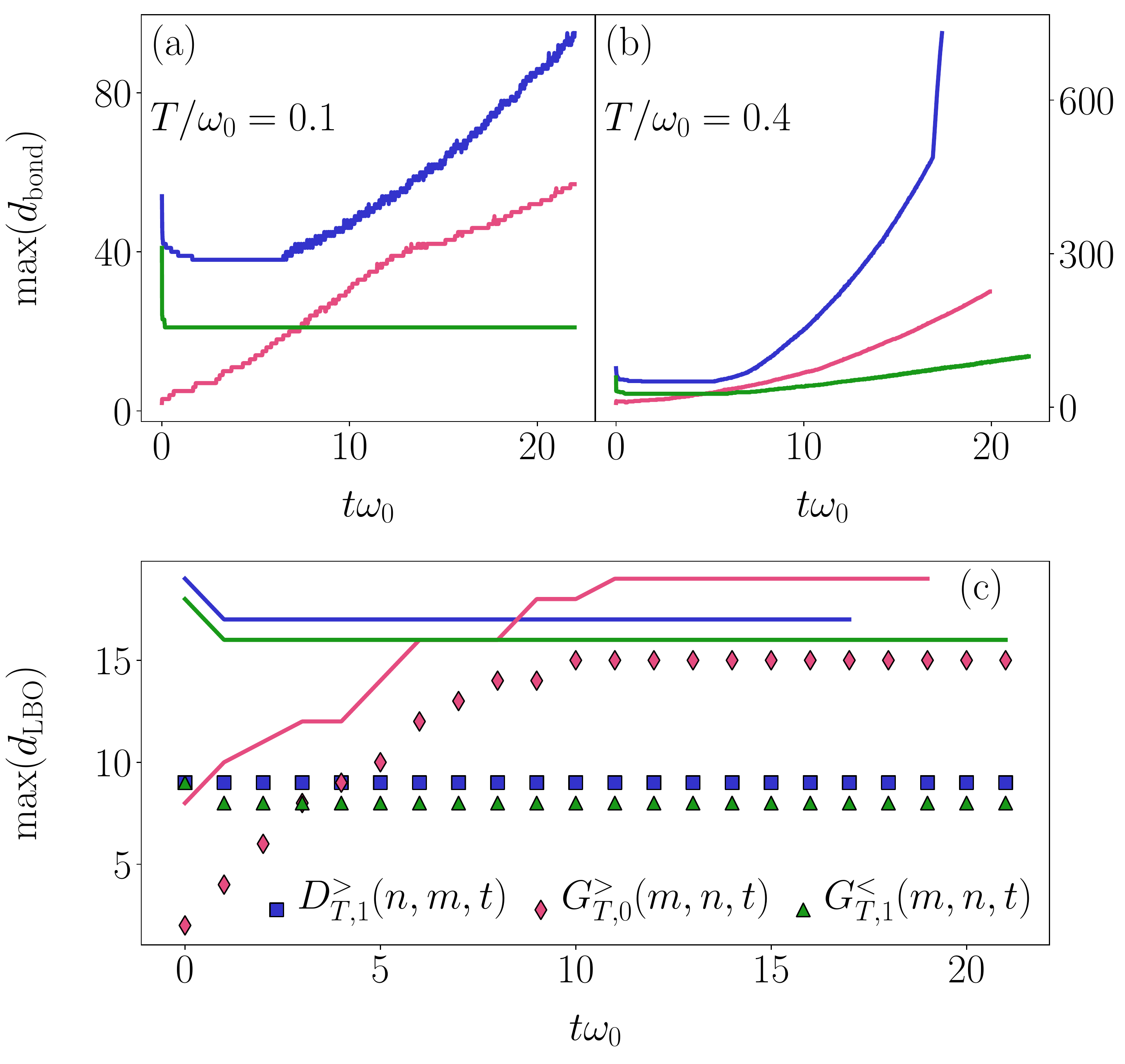}
\caption{(a) Maximum bond dimensions used in the MPS representations of $D_{T,1}^>(m,n, t)$ (blue), $G_{T,0}^>(m,n, t)$ (red) and $G_{T,1}^<(m,n, t) $ (green) [see Eqs.~\eqref{eq:def_corelD}, \eqref{eq:def_corel0}, \eqref{eq:def_corel1}] at $T/\omega_0=0.1$. The parameters are  $\gamma / \omega_0=\sqrt{2},t_0 / \omega_0=1, L=21,M=20$ and fixed $m=n=11$. (b) Same as in (a) but at $T/\omega_0=0.4$.  (c) Maximum local optimal basis dimension for $T/\omega_0=0.1$ (symbols) and $T/\omega_0=0.4$ (solid lines). We only show every 100th point for the symbols in (c) for clarity.}
\label{fig:lbobd}
\end{figure}
 \begin{figure}[t]
    \centering
\includegraphics[width=0.99\columnwidth]{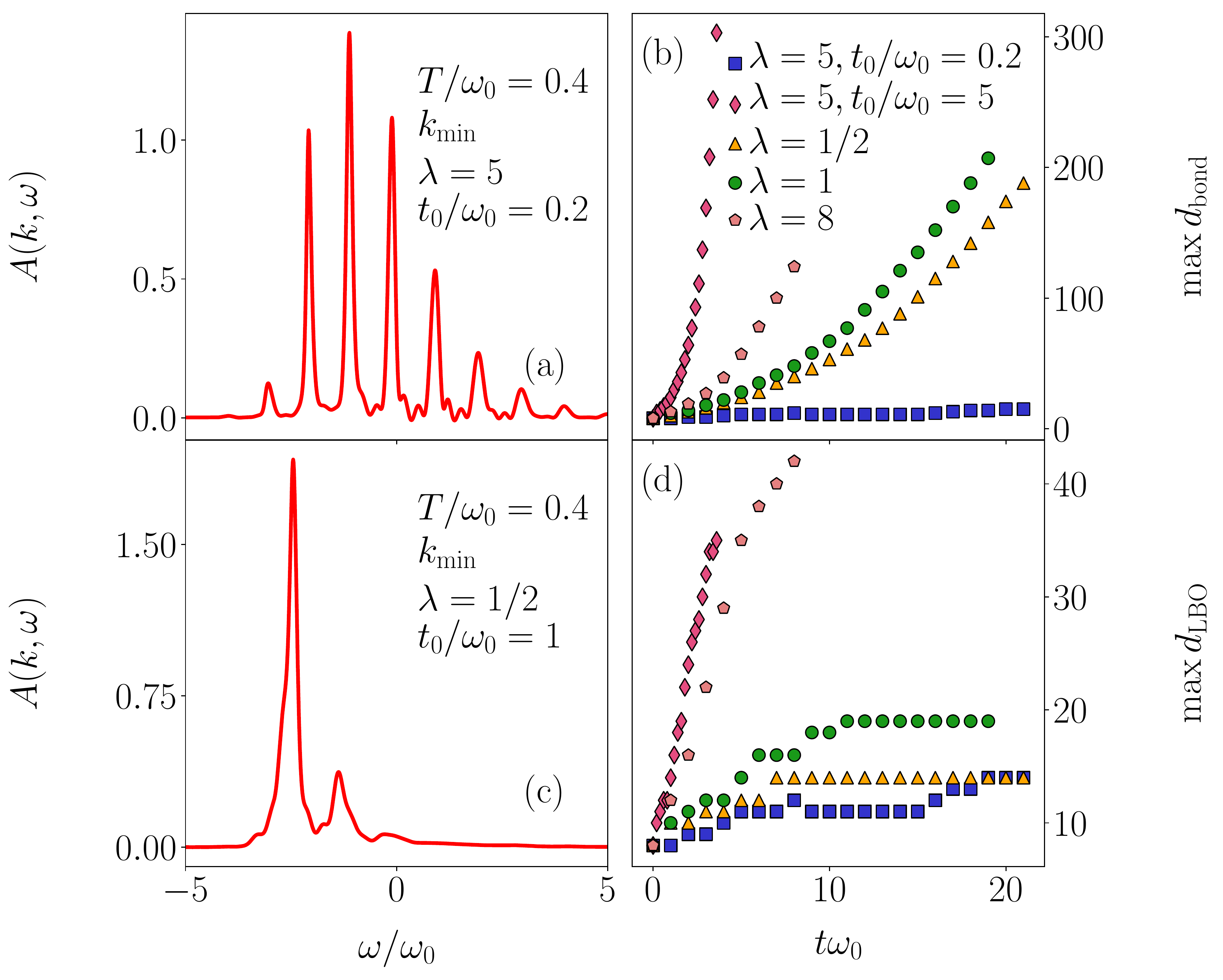}
\caption{(a) Electron spectral function [see Eq.~\eqref{eq:def_ferspec}], for $\gamma/\omega_0=\sqrt{2}, t_0 / \omega_0=0.2, L=21, M=20,\eta=0.05, T/\omega_0=0.4 ,k=\pi /(L+1)$ and $t_{\rm max}\omega_0=18.0$. (c) Same as (a) but with $\gamma/\omega_0=1, t_0 / \omega_0=1 $. (b) Maximum bond dimension of $G_{T,0}^>(m,n, t)$ for different parameters at $T/\omega_0=0.4$. (d) Maximum local dimension of $G_{T,0}^>(m,n, t)$ for different parameters at $T/\omega_0=0.4$. In (b) and (d), we only show every 100th point for clarity. } 
\label{fig:diffparams}
\end{figure}
In Fig.~\ref{fig:lbobd}, we show the maximum bond dimension [Fig.~\ref{fig:lbobd}(a) and Fig.~\ref{fig:lbobd}(b)] and maximum local optimal basis dimension [Fig.~\ref{fig:lbobd}(c)] as a function of time. The bond dimension is clearly dependent on the temperature and on the specific Greens's function. It increases a lot faster for the $G_{T,0}^>(t,m,n)$ (red) and $D^>_{T,1}(t,m,n)$ (blue). For these Greens's functions, the real-time evolution is done with an electron in the physical system which causes the large increase in the bond dimension. One also sees that as  temperature is increased, the computations become much more costly. This is especially true for the phonon Green's function $D^>_{T,1}(t,m,n)$.  The drop at $t\omega_0=0$ comes from the fact that the imaginary-time evolution is carried out with $\rho_{\rm LBO}=\rho_{\rm bond}=10^{-9}$, whereas the real-time evolution is done with $ \rho_{\rm LBO}=\rho_{\rm bond}=10^{-8}$. This leads to some states getting truncated away right at the beginning. This is not the case for the red curve that shows $G_{T,0}^>(m,n, t)$. There, the insertion of the electron into the system directly leads to a much larger bond dimension.
We also observe that the maximum dimension of the local optimal basis remains approximately constant during the real-time evolution for the Green's functions in the one-electron sector. The dimension clearly increases for larger temperature  [$T/\omega_0 =0.1$ (symbols) and $T/\omega_0 =0.4$ (solid lines)] but it is, in both cases, clearly beneficial. This does, of course, not imply that the modes in the optimal basis remain the same. When the Green's function is calculated in the zero-electron sector, inserting the electron clearly leads to an increase in $\max(d_{\rm LBO})$.

We want to explore the performance of our method away from the intermediate coupling regime. The strength and purpose of our DMRG method is to access the cross-over regime, whereas perturbation theory~\cite{langfirsov} can be used to address the small hopping limit. This is illustrated in Fig.~\ref{fig:diffparams}, where we show the results for different choices of $\gamma$ and $\omega_0$.

Figure~\ref{fig:diffparams}(a) shows the electron spectral function [see Eq.~\eqref{eq:def_ferspec}] for $t_0 / \omega_0 = 0.2$ and $\gamma / \omega_0=\sqrt{2}$. This is close to the atomic limit and is in good agreement with the single-site spectral function presented in Fig.~\ref{fig:singlepec}. In Fig.~\ref{fig:diffparams}(c), the same quantity is shown for $\gamma / \omega_0=1$ and $t_0 / \omega_0 = 1$.

The limiting factors for the performance are displayed in Figs.~\ref{fig:diffparams}(b) and (d). Figure~\ref{fig:diffparams}(b) shows the maximum bond dimension of the matrix-product state for different parameters as a function of time. For both a large coupling and small frequencies, the bond dimension grows more rapidly than for the parameters used in the main text. This makes the real-time evolution significantly more difficult. Furthermore,  Fig.~\ref{fig:diffparams}(d) shows the maximum LBO dimension. This quantity also increases more rapidly in both previously mentioned cases, rendering the use of LBO more costly. A sufficient number of bare phonons $M$ is clearly not included in those simulations. To make accurate computations in these parameter regimes, the convergence with increased $M$ would have to be monitored. This can be done adaptively in DMRG-LBO simulations as was demonstrated by Brockt in Ref.~\cite{brockt_phd}.
 
  Intuitively, one would think that LBO works better in the strong-coupling limit, since in the Lang-Firsov limit, one should be able to describe the system with  only two local states at $T/\omega_0=0$. As expected, this is the case for $t_0 /\omega_0=0.2$. In Fig.~\ref{fig:diffparams}(d), we see that we need $\sim O(10)$ states at later times in this regime. Further, the single-site polaron ground state has a phonon occupation $N_{\rm ph}= \gamma^2/ \omega_0^2$. This would, for example, give $N_{\rm ph}=16$ for the $\lambda=8$ curve in Fig.~\ref{fig:diffparams}(d), such that the optimal basis for the distribution is out of reach for the $M=20$ bare-phonon truncation used here.
   
 For the data shown here, we start the time-evolution in the trivially obtained zero-electron state. In the process of generating the one-electron state with imaginary time-evolution, the particle might jump to the ancilla sites at low temperatures in extreme parameter regimes. This can be overcome with standard solutions, see Refs.~\cite{barthel_16,nocera_16}. One possibility is to  generate matrix-product states with particle-number conservation in the physical and ancilla system separately.

\section{Error of the electron spectral function} \label{sec:app0}
\begin{figure}[t]
    \centering
\includegraphics[width=0.99\columnwidth]{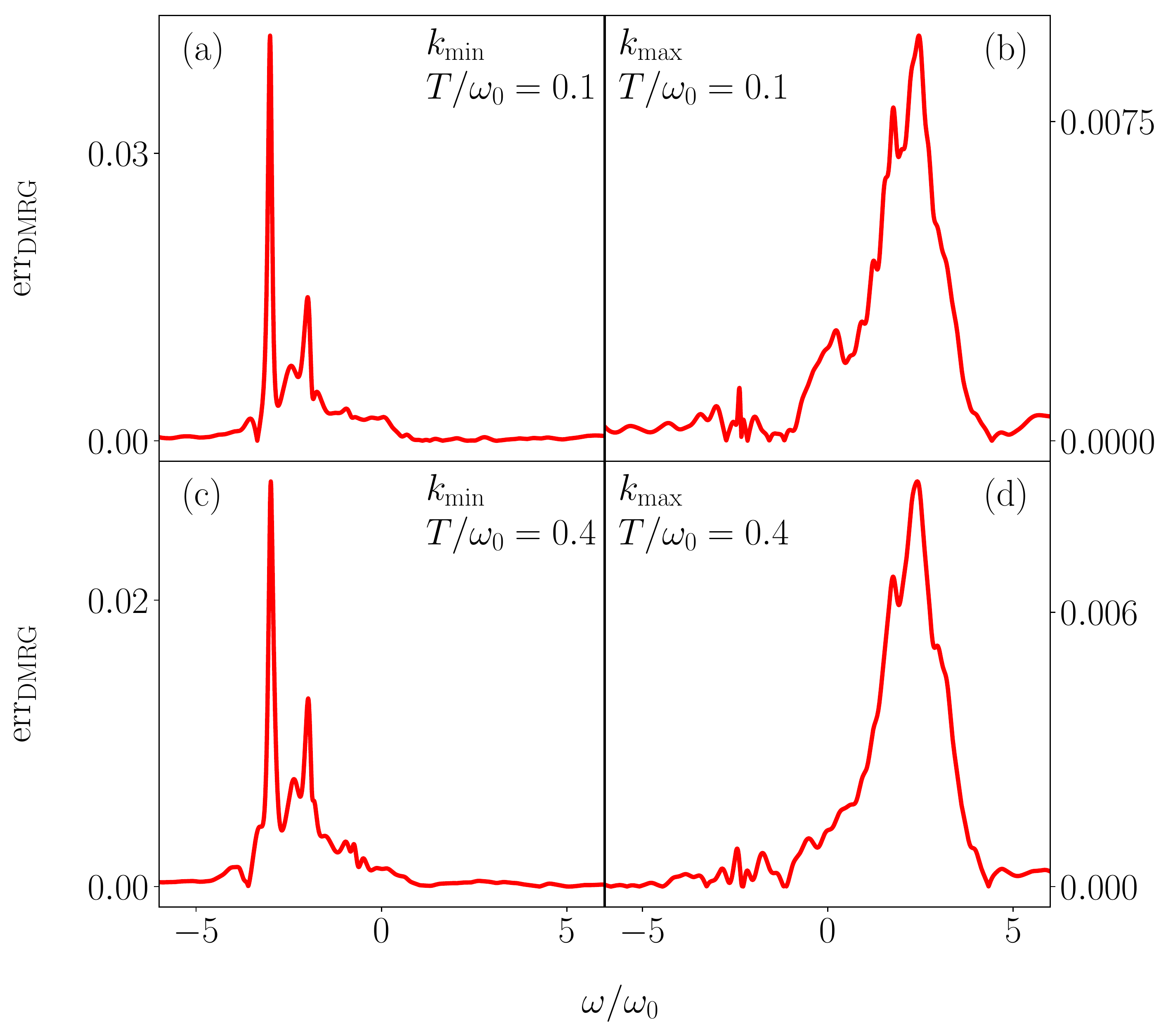}
\caption{
Difference between DMRG data at different system sizes defined in Eq. \eqref{eq:def_errdmrg}. 
The parameters are the same as in Fig.~\ref{fig:G0specdL}. }
\label{fig:diffdLdmrg}
\end{figure}

In Fig.~\ref{fig:diffdLdmrg}, we show the difference between the integrated [see, Eq.~\eqref{eq:def_I}] DMRG data with $L=21$ and $L=101$
\begin{equation}
\label{eq:def_errdmrg}
\text{err}_{\rm DMRG}=\frac{\abs{I(k, \omega)_{L=21}-I(k, \omega)_{L=101}}}{\max{\{I(k, \omega)_{L=21}}\}}.
\end{equation}
The data is obtained with different Fourier transformations, see Sec.~\ref{subsubsec:espec} for details.

\section{Moments} \label{sec:app2}
\begin{figure}[!]
    \centering
\includegraphics[width=0.99\columnwidth]{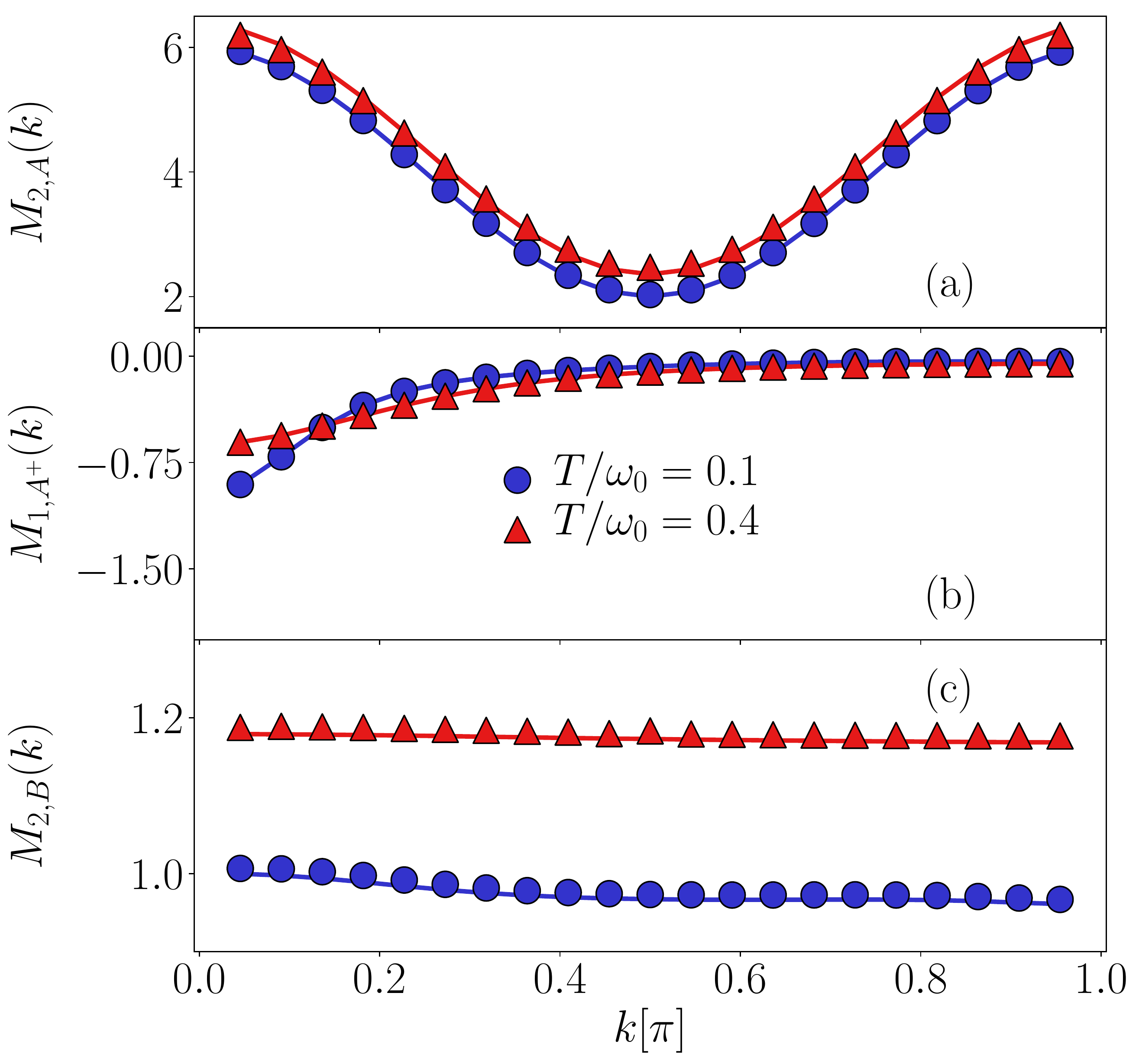}
\caption{Temperature dependent moments. The symbols are calculated by numerically integrating the moments as in Eq.~\eqref{eq:def_moments} and the solid lines correspond to the thermal expectation values thereof.  For all data, we use a Gaussian regularization and $\eta=0.05/(6\pi)$. (a) Second moment of the electron spectral function for the same parameters as in Fig.~\ref{fig:G0dens}.  (b) First moment of the electron emission spectrum for the same parameters as in Fig.~\ref{fig:AEMdens}. (c) Second moment of the phonon spectral function for the same parameters as in Fig.~\ref{fig:Bspecdense}. }
\label{fig:moms}
\end{figure}
To validate that the method captures the correct finite-temperature behaviour we compute the first temperature-dependent moment for each spectral function. The moments are defined for the corresponding spectral function [here only shown for $A(k, \omega)$] as 
\begin{equation}
\label{eq:def_moments}
M_{m,A}(k)=\int_{-\infty}^{\infty} \omega^m A(k,  \omega) d \omega \,.
\end{equation}
For $A(k, \omega)$, the first two moments (see Refs.~\cite{Kornilovitch_2002,goodvin06,bonca2019}) become
\begin{align}
\begin{split} \label{eq:def_momAper1}
M_{1, A}(k)=\epsilon(k),
\end{split} \\
\begin{split}
\label{eq:def_momAper2}
 M_{2,A}=\epsilon^2(k) + \gamma^2(2 n_{\rm ph}+1),
 \end{split}
\end{align}
where $n_{\rm ph}=1/(\exp(\omega_0/T)-1)$, $\epsilon(k)=-2t_0\cos(k)$ with the quasi momenta for open-boundary conditions used in this paper.
For $A^{+}(k, \omega) $, the first moment is already temperature dependent
\begin{multline}
\label{eq:def_momAp}
M_{1,A^+}(k)=\frac{2}{L+1} \sum\limits_{i, j} \sin(ki)\sin(kj) \\ 
\times \expval*{\hat c^{\dagger}_j \hat c_{i}(\epsilon(k)+\gamma  \hat X_i)}_T,
\end{multline}
and for $B(k, \omega) $ we obtain 
\begin{align}
\begin{split}\label{eq:def_momB1}
M_{1,B}(k)=\omega_0 , 
\end{split} \\
\begin{split} \label{eq:def_momB2}
M_{2,B}(k)=\omega_0\frac{2}{L+1}\sum\limits_{i,j}\sin(ik)\sin(jk) \\ \times\expval*{\omega_0(\hat X_i \hat X_j)+2\gamma \hat n_i \hat X_j}_T.
\end{split}
\end{align}
The results for the temperature-dependent moments are shown in Fig.~\ref{fig:moms}. We see that they can be calculated quite accurately with our method.  The mean differences for both temperatures are of the order $O(10^{-5})$ for all the first moments and  $O(10^{-2})$ for the second moments.
In contrast to the rest of the paper, we  here use a Gaussian regularization for the spectral function. The moments show a dependence on the regularization parameter $\eta $. One must find a compromise between allowing for unphysical oscillations in the spectral function and the accuracy of the moments. We choose $\eta=0.05/(6 \pi) $. For the second moments, we further limit the integration to  $\omega_{\min}\approx-10 \omega_0 < \omega <\omega_{\rm max}\approx 10 \omega_0$.
We found that the results for the Gaussian regularization are much  more robust against changes in $\omega_{\min}$ and $\omega_{\max}$ than the Lorenzian regularization.

\bibliographystyle{biblev1}
\bibliography{references}

\end{document}